\documentclass[letterpaper,english,reprint,pre,longbibliography,aps,superscriptaddress]{revtex4-1}
\usepackage[T1]{fontenc}
\usepackage[latin9]{inputenc}
\usepackage{babel}
\usepackage{verbatim}
\usepackage{mathtools}
\usepackage{bm}
\usepackage{amsmath}
\usepackage{amssymb}
\usepackage{graphicx}
\usepackage{color}
\usepackage[unicode=true,
 bookmarks=true,bookmarksnumbered=false,bookmarksopen=false,
 breaklinks=false,pdfborder={0 0 1},backref=false,colorlinks=false]
 {hyperref}

\usepackage{ulem}
\usepackage{bbold}

\makeatletter

\pdfpageheight\paperheight
\pdfpagewidth\paperwidth

\newcommand{\PtubeDLdot}{\dot{\tilde{P}}^{\,\trajvec}_R}
\newcommand{\xveczeroDL}{\xzeroDLvec}
\newcommand{\Pzero}{{P}_i}
\newcommand{\Stub}{\mathcal{S}}
\newcommand{\SumDL}{\tilde{\mathcal{S}}}
\newcommand{\epsilonDL}{\tilde{\epsilon}}
\newcommand{\yDL}{\tilde{y}}
\newcommand{\aDL}{\tilde{a}}
\newcommand{\partialDL}{\tilde{\nabla}}
\newcommand{\avec}{\bm{a}}
\newcommand{\avecDL}{\tilde{\bm{a}}}
\newcommand{\Wvec}{\bm{W}}
\newcommand{\bmat}{\underline{\bm{b}}}
\newcommand{\Dmat}{\underline{\bm{D}}}
\newcommand{\DmatDL}{\tilde{\Dmat}}
\newcommand{\traj}{\varphi}

\newcommand{\trajTwo}{\psi}
\newcommand{\LagrangianOnsagerMachlup}{\mathcal{L}_{\mathrm{OM}}}
\newcommand{\SOM}{\mathcal{S}_{\mathrm{OM}}}
\newcommand{\Dexp}{\tilde{\mathcal{D}}}
\newcommand{\tDL}{\tilde{t}}
\newcommand{\tzeroDL}{\tilde{t}_i}

\newcommand{\tfinal}{{t}_f}
\newcommand{\tinitial}{0}
\newcommand{\tinitialDL}{\tzeroDL}
\newcommand{\BR}{B_R^{\trajvec}}
\newcommand{\BDL}{\tilde{B}}
\newcommand{\tDLdummy}{\tilde{t}\,'}
\newcommand{\xDL}{\tilde{x}}

\newcommand{\xzeroDLvec}{\tilde{\bm{x}}_i}

\newcommand{\vecX}{\bm{X}} 
\newcommand{\Xvec}{\vecX}
\newcommand{\vecx}{\bm{x}} 
\newcommand{\xvec}{\vecx} 
\newcommand{\vectraj}{\bm{\traj}}
\newcommand{\trajvec}{\vectraj}
\newcommand{\vectrajdot}{\dot{\vectraj}}
\newcommand{\grad}{\bm{\nabla}}
\newcommand{\vectrajTwo}{\bm{\trajTwo}}

\newcommand{\trajTwovec}{\vectrajTwo}
\newcommand{\vecxDL}{\tilde{\bm{x}}}
\newcommand{\xvecDL}{\vecxDL}
\newcommand{\xDLvec}{\vecxDL}
\newcommand{\vectrajDL}{\tilde{\bm{\traj}}}
\newcommand{\vectrajdotDL}{\dot{\tilde{\vectraj}}}
\newcommand{\alphavec}{\bm{\alpha}}
\newcommand{\IntDL}{\tilde{\mathcal{I}}}
\newcommand{\DDL}{\tilde{D}}
\newcommand{\trajdotDL}{\dot{\tilde{\varphi}}}
\newcommand{\reqDL}{\tilde{\rho}_{\mathrm{ss}}}
\newcommand{\TubularEnsemble}{\mathcal{X}_R^{\trajvec}}
\newcommand{\Ptube}{P^{\,\bm{\varphi}}_R}
\newcommand{\Ptubedot}{\dot{P}^{\,\bm{\varphi}}_R}
\newcommand{\PtubeTwo}{P^{\,\bm{\psi}}_R}
\newcommand{\PtubeDL}{\tilde{P}^{\,\bm{\varphi}}_R}
\newcommand{\PtubedotDL}{\PtubeDLdot}
\newcommand{\setbar}{\bigl\vert}
\newcommand{\aexit}{\alpha_{R}^{\trajvec}}
\newcommand{\aexitDL}{\tilde{\alpha}}
\newcommand{\aexitfreeDL}{\tilde{\alpha}_{\mathrm{free}}}
\newcommand{\aexitfree}{{\alpha}_{\mathrm{free}}}
\newcommand{\Pinitial}{\Pzero}
\newcommand{\td}{\tau_D}

\newcommand{\trelDL}{\tilde{\tau}_{\mathrm{rel}}}

\newcommand{\LFPDLapp}{\tilde{\mathcal{F}}_{\mathrm{app}}}
\newcommand{\evn}{\tilde{\lambda}_{n}}
\newcommand{\evone}{\tilde{\lambda}_{1}}
\newcommand{\efn}{\tilde{\rho}_n}
\newcommand{\efnk}{\efn^{(k)}}
\newcommand{\efm}{\tilde{\rho}_m}
\newcommand{\efndot}{\dot{\tilde{\rho}}_n}
\newcommand{\efmdot}{\dot{\tilde{\rho}}_m}
\newcommand{\efone}{\tilde{\rho}_1}
\newcommand{\efonedot}{\dot{\tilde{\rho}}_1}
\newcommand{\Qs}{\tilde{Q}_{n,\mathrm{s}}}
\newcommand{\Qc}{\tilde{Q}_{n,\mathrm{c}}}
\newcommand{\Qsk}{\tilde{Q}_{n,\mathrm{s}}^{(k)}}
\newcommand{\Qck}{\tilde{Q}_{n,\mathrm{c}}^{(k)}}
\newcommand{\Ln}{\tilde{\Lambda}_n}
\newcommand{\Lone}{\tilde{\Lambda}_1}
\newcommand{\Lm}{\tilde{\Lambda}_m}
\newcommand{\Lmn}{\Delta \tilde{\Lambda}_{mn}}
\newcommand{\Lmone}{\Delta \tilde{\Lambda}_{m1}}
\newcommand{\Lnone}{\Delta \tilde{\Lambda}_{n1}}
\newcommand{\Lnm}{\Delta \tilde{\Lambda}_{nm}}
\newcommand{\EDL}{\tilde{E}}
\newcommand{\SR}{S_{\mathrm{r}}}
\newcommand{\xl}{x_-}
\newcommand{\xr}{x_+}

\makeatother

\begin{document}
\title{Sojourn probabilities in tubes and pathwise irreversibility for It{\^o} processes}
\author{Julian Kappler}
\email{jkappler@posteo.de}
\affiliation{Department of Applied Mathematics and Theoretical Physics, Centre for Mathematical Sciences, University of Cambridge, Wilberforce Road, Cambridge CB3 0WA, United Kingdom}
\affiliation{Arnold Sommerfeld Center for Theoretical Physics (ASC), Department of Physics, Ludwig-Maximilians Universit\"at M\"unchen, Theresienstra{\ss}e 37, D-80333 Munich, Germany}
\author{Michael E.~Cates}
\affiliation{Department of Applied Mathematics and Theoretical Physics, Centre for Mathematical Sciences, University of Cambridge, Wilberforce Road, Cambridge CB3 0WA, United Kingdom}
\author{Ronojoy Adhikari}
\affiliation{Department of Applied Mathematics and Theoretical Physics, Centre for Mathematical Sciences, University of Cambridge, Wilberforce Road, Cambridge CB3 0WA, United Kingdom}
\date{\today}
\begin{abstract}
 The sojourn probability of an It{\^o} diffusion process, that is its probability to remain in the tubular neighborhood of a smooth path, is a central quantity in the study of path probabilities.
For $N$-dimensional It{\^o} processes with
a state-dependent full-rank diffusion tensor,
we derive a general expression for the sojourn probability in tubes whose radii are small but finite, 
and fixed by the metric of the ambient Euclidean space.
The central quantity in our study is the exit rate at which trajectories leave the tube for the first time. This has an interpretation as a Lagrangian and can be measured directly in experiment, 
unlike previously defined sojourn probabilities (involving variable tube radius or shape) which 
depend on prior knowledge of the state-dependent diffusivity.
We find that while in the limit of vanishing tube radius the ratio of sojourn probabilities for a pair of distinct paths is in general divergent, the same for a path and its time-reversal is always convergent and finite. 
This provides, in turn, a pathwise definition of irreversibility for It{\^o} processes that is agnostic to the state-dependence of the diffusion tensor.
For one-dimensional systems we derive an explicit expression for our Lagrangian
in terms of the drift and diffusivity,
and find that our Lagrangian differs from all previously
reported multiplicative-noise generalizations of the
Onsager-Machlup Lagrangian.
 We confirm our result by comparing to numerical simulations for a one-dimensional diffusion process with state-dependent diffusivity,
 and relate our theory to the classical Stratonovich  Lagrangian for multiplicative noise.
For one-dimensional systems, we furthermore discuss under which conditions the vanishing-radius limiting ratio of sojourn probabilities for a
pair of forward and backward paths recovers the pathwise entropy production found in the literature.
Finally, 
we demonstrate for our
one-dimensional example system that the most probable tube for a barrier crossing depends sensitively on the tube radius, 
and hence on the tolerated amount of fluctuations around the smooth reference path.
\end{abstract}
\maketitle

\section{Introduction}

The overdamped Langevin equation is a fundamental model
for diffusive stochastic dynamics with a wide range of applications 
ranging from chemical physics \cite{kampen_stochastic_2007}, 
to ecology \cite{nolting_balls_2016} and finance \cite{friz_large_2015,oksendal_stochastic_2007}.
For the special case of additive noise, meaning that the noise term in
 the Langevin equation does not depend on the current state 
 of the system,
relative path probabilities are well-characterized \cite{onsager_fluctuations_1953,
stratonovich_probability_1971,
horsthemke_onsager-machlup_1975,
graham_path_1977,
durr_onsager-machlup_1978,ito_probabilistic_1978,
wissel_manifolds_1979,
langouche_functional_1979,
dekker_path_1980,
williams_probability_1981,
fujita_onsager-machlup_1982,
ikeda_stochastic_1989,
adib_stochastic_2008,
weber_master_2017,
cugliandolo_building_2019}.
In particular, one experimentally relevant approach to path probabilities is via
 the sojourn probability, i.e.~the probability
that a stochastic trajectory always remains within a 
ball of constant finite radius and with moving center given by
a
 twice continuously differentiable
reference path.
Since for diffusive dynamics any given individual path is
 observed with vanishing probability, the sojourn probability 
becomes zero in the limit of vanishing radius.
However, for any given pair of paths 
the  ratio of sojourn probabilities has a finite
 vanishing-radius limit,
so that this limiting ratio can be used to define a ratio of path probabilities.
It was Stratonovich \cite{stratonovich_probability_1971} who first showed
 that the limiting ratio of sojourn probabilities is characterized by the 
Onsager-Machlup (OM) stochastic action functional \cite{stratonovich_probability_1971,
durr_onsager-machlup_1978,ito_probabilistic_1978,
williams_probability_1981,ikeda_stochastic_1989,
kappler_stochastic_2020}.
An advantage of considering path probabilities
as limiting ratios of sojourn probabilities is the connection to physical observables;
for any finite tube radius the sojourn probability is positive,
and hence an experimental observable. 
Indeed, the mathematical theory of Stratonovich \cite{stratonovich_probability_1971} was recently
confirmed experimentally, by extrapolating measured ratios of finite-radius sojourn probabilities
to the single-path limit of vanishing radius \cite{gladrow_experimental_2021}.

While for additive noise the situation is well-understood, 
 the situation is not as simple
for multiplicative noise, where the noise strength in the Langevin equation 
depends on
 the current state of the system \cite{kampen_stochastic_2007}.
For state-dependent noise, the limiting ratio of 
tube probabilities
in general does not  yield meaningful results,
and defining a probability density relative to a quasi translation invariant measure
 on the space of all paths is not possible \cite{durr_onsager-machlup_1978}.
Still, starting with the work of Freidlin and Wentzel \cite{ventsel_small_1970} (who focused
on the weak-noise limit, as discussed more in the conclusions),
and Stratonovich \cite{stratonovich_probability_1971}, 
there have been several attempts
to quantify relative path probabilities also in systems with multiplicative noise
\cite{horsthemke_onsager-machlup_1975,ito_probabilistic_1978}.
These works
can broadly be 
classified into two approaches.
One approach is to define the tubular neighborhood using the metric induced by the diffusion tensor
of the stochastic dynamics \cite{ventsel_small_1970}.
The tube is then a moving ellipsoid in $\mathbb{R}^N$,
whose principal axes may vary along the reference path, which is the
 geometric center of mass of the ellipsoid.
Another approach is to locally introduce a new coordinate system, 
relative to which the diffusivity
is state-independent \cite{stratonovich_probability_1971,
 ito_probabilistic_1978}. 
In this new coordinate system, the established theory for
 additive noise
can be applied.
However,
since the relation between the two sets of  
coordinates is nonlinear,
in the original coordinates 
the tube can have an in principle arbitrary geometrical shape.
In both approaches mentioned here the tube is therefore not a moving ball in Euclidean space, but an
ellipsoid or more general geometrical shape.
\textcolor{black}{
As we show further below in Sect.~\ref {sec:results_for_one_dimensional_system} for one-dimensional systems, 
these two approaches lead to different small-radius stochastic action Lagrangians.}
Importantly, in both cases the diffusivity tensor associated with the underlying stochastic dynamics
 needs to be known to even
construct the tube.
\textcolor{black}{
This severely limits the experimental relevance of these sojourn probabilities,
as in an experimental system it is natural to consider a constant-radius tube with respect
to the Euclidean metric of the ambient space (and knowledge of the diffusivity
 profile might not even be available).
Indeed, the recent experimental measurement of the OM stochastic action for additive noise 
employed constant-radius tubes \cite{gladrow_experimental_2021}.
}
Hitherto no theory existed
for the sojourn probability of a constant-radius scenario for multiplicative noise.

\begin{figure}[ht]
\centering \includegraphics[width=1\columnwidth]{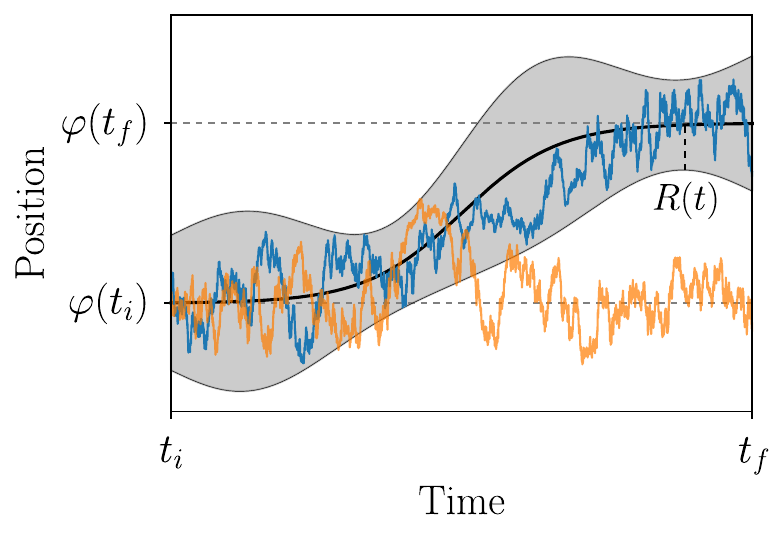} 
\caption{ \label{fig:tubular_ensemble} 
Illustration of the tubular neighborhood of a reference path.
The black solid line depicts a reference path $\traj(t)$ that starts at $\traj(t_i)$ and 
ends at $\traj(t_f)$, as indicated by two horizontal gray dashed lines. 
The gray shaded region around $\traj$ depicts a tube with time-dependent radius $R(t)$
around the reference path, the instantaneous tube radius is indicated by a vertical dashed line.
The blue solid line is a realization of one-dimensional
 Langevin dynamics with multiplicative noise which remains
within the tube until the final time, 
the orange solid line represents a realization that leaves the tube before the final time $t_f$.
}
\end{figure}

We here fill this gap, by providing a comprehensive theory of 
tube probabilities 
for diffusive dynamics with state-dependent diffusivity.
We achieve this by establishing an
 expression for the sojourn probability for tubes with small-but-finite radius, which
may vary along the reference path, as illustrated in Fig.~\ref{fig:tubular_ensemble}.
The central quantity in our theory is the exit rate at which trajectories first leave the tube,
which can be interpreted as a generalization of the stochastic action Lagrangian.
We present a series expansion of this exit rate in powers of the 
time-dependent tube radius.
This power series generalizes a previous result for the tubular exit rate
 for additive noise \cite{kappler_stochastic_2020},
to which it reduces if the diffusion tensor of the stochastic dynamics is 
state-independent and isotropic, and if additionally the tube radius is independent of
time.

Our theory for finite-radius tubes
 leads to a 
 physical picture as to why the definition of ratios of path probabilities
is not straightforward for state-dependent diffusivity.
Namely, 
because diffusive stochastic dynamics with a full-rank diffusivity tensor is at short
 length- and time scales dominated by
the random noise, as compared to the drift, 
the sojourn probability is for asymptotically small radius dominated by the 
diffusivity \cite{kappler_stochastic_2020}.
For additive noise this dominant term leads to a path-independent factor in the sojourn probability, 
and so is irrelevant for ratios of path probabilities \cite{kappler_stochastic_2020}.
However, this is not the case for state-dependent diffusivity, 
and is the reason why limiting ratios of tube probabilities are in general either zero or divergent.
As we discuss further below, finite limiting ratios of sojourn probabilities
are only obtained if the tube radius is fine-tuned in such a way that the dominating 
short-time noise contributions cancel.

While limiting ratios of sojourn probabilities in general
do not yield meaningful results, we show that the ratio for a pair of
 forward and time-reverse reference path
 is always finite. 
This provides  a pathwise definition of irreversibility for It{\^o} processes that is
 agnostic to the state-dependence of the diffusion tensor,
 and which is related to classical measures of irreversibility and 
 entropy production \cite{seifert_entropy_2005,bo_functionals_2019,kappler_measurement_2022}.

For the special case of a one-dimensional system, we derive explicit
expressions for the exit rate describing the sojourn probability, 
 and discuss several choices for the time-dependent tube radius.
 In particular we present an explicit formula for the sojourn probability of a constant-radius
 tube.
We validate our theory by comparing to numerical simulations, 
and  
discuss explicitly the relation of the Stratonovich Lagrangian \cite{stratonovich_probability_1971}
to tubular exit rates.
In the context of barrier crossing for a one-dimensional system,
 we furthermore show that the most probable
tube depends sensitively on both the details of the time-dependence of the radius, and the size of the tube.

In our accompanying paper Ref.~\cite{accompanying_paper_short} we compare the theoretical sojourn probabilities
derived here also to experimental results,  and discuss the radius
 dependence of the most probable tube based on measured time series.
The present work thus
provides an experimentally accessible
 approach to quantifying and generalizing path probabilities 
for systems with state-dependent diffusivity.

The main part of this paper is structured as follows.
In Sect.~\ref{sec:Theory}, we first present our general
 theory of sojourn probabilities for $N$-dimensional Langevin dynamics 
 with arbitrary state-dependent full-rank
diffusion matrix.
In Sect.~\ref{sec:results_for_one_dimensional_system} we consider
one-dimensional
systems, $N=1$, for which we provide explicit expressions for the exit rate
in terms of the diffusivity and drift,
 discuss several possible choices for $R(t)$,
and relate the Stratonovich Lagrangian to
tubular exit rates.
We furthermore discuss how the most probable tube connecting an
initial and a final state can depend on the tube radius.
We conclude in Sect.~\ref{sec:summary_and_conclusions}, where we summarize our findings and 
 discuss their further implications.

\section{$N$-dimensional theory}

\label{sec:Theory}

We now present our general results for $N$-dimensional diffusive dynamics.
For an $N$-dimensional coordinate $\Xvec_{t}\equiv\Xvec(t)\equiv(X_{1}(t),...,X_{N}(t))$,
we consider the It\^{o}-Langevin equation
 given by \cite{kampen_stochastic_2007,gardiner_stochastic_2009}
\begin{equation}
\mathrm{d}{\vecX}_{t}= \avec(\vecX_{t},t)\,\mathrm{d}t+\bmat (\vecX_t,t) \,\mathrm{d}\Wvec_{t},\label{eq:Ito_process_definition}
\end{equation}
where $d\Wvec_{t}$ is the increment of the $N$-dimensional Wiener process,
 $\avec$ is the drift, and $\bmat$ is the noise matrix,
with components
$a_i \equiv a_i(\xvec,t)$, $b_{ij} \equiv b_{ij}(\xvec,t)$.
While we interpret Eq.~\eqref{eq:Ito_process_definition} in the It{\^o}
sense, results for other conventions are obtained 
from our results by 
modifying the drift term $\avec$ appropriately \cite{gardiner_stochastic_2009}.
The stochastic dynamics defined by Eq.~\eqref{eq:Ito_process_definition} can equivalently
be described via the Fokker-Planck equation (FPE) \cite{gardiner_stochastic_2009}
\begin{align}
\partial_{t}P&=
-\nabla_i \left(
a_i 
P\right)
\label{eq:ND_FP_eq_intro} 
+\nabla_i \nabla_j \left(D_{ij} P\right),
\end{align}
where $P \equiv P(\xvec,t\mid \xvec_i, t_i)$ is the transition probability density for finding a particle 
at position $\xvec$ and time $t$ after it has started at $\xvec_i$ at time $t_i$,
by $\nabla_i \equiv \partial/\partial x_i \equiv \partial_i$ we denote
 the partial derivative in the $x_i$-direction,
and the components
of the symmetric diffusion tensor $\Dmat$ are given by
$D_{ij}(\xvec,t) \equiv b_{ik}(\xvec,t) b_{jk}(\xvec,t)/2$,
where 
 we use the Einstein sum convention for repeated indices.
 We assume that the diffusivity tensor $\Dmat$ is a matrix of full rank, which physically means
 that the noise directly acts on all $N$ degrees of freedom.
To denote time derivates, we use the notation $\partial_t P$ and $\dot{P}$
 interchangeably.

\subsection{The tubular ensemble and sojourn probabilities}
\label{sec:the_tubular_ensemble}

We consider the tubular ensemble, which consists of all realizations of
the Langevin Eq.~\eqref{eq:Ito_process_definition} that remain within a time-dependent distance $R(t)$
of a continuous reference path $\trajvec(t)$ until time $\tfinal$
\cite{ventsel_small_1970,
stratonovich_probability_1971,
durr_onsager-machlup_1978,
williams_probability_1981,
horsthemke_onsager-machlup_1975,
zeitouni_onsager-machlup_1989,
ito_probabilistic_1978,
fujita_onsager-machlup_1982,
ikeda_stochastic_1989},
\begin{align}
\TubularEnsemble(\tfinal) &\equiv\left\{ \,\Xvec\,\,\setbar\,\,||\Xvec_{s}-\trajvec(s)||<R(s)~~\forall~s\in[0,\tfinal]\,\right\},
\end{align}
where in principle any norm $||~||$ on $\mathbb{R}^N$ can be used to quantify distances.
We use the name tubular ensemble
 for $\TubularEnsemble$ because a ball
  with center $\trajvec(t)$ and
 radius $R(t)$
 is a tube in spacetime $(\xvec,t)$, as illustrated in Fig.~\ref{fig:tubular_ensemble}.

The sojourn probability
\begin{align}
\Ptube(t) &
\label{eq:sojourn_probability}
\equiv P\left(\,\Xvec\in\TubularEnsemble(t)\,\setbar\,\Xvec_{\tinitial}\sim\Pinitial\,\right)
\end{align}
is the probability that a stochastic trajectory $\Xvec$ remains within
the tubular neighborhood around the reference path $\trajvec$ until time $t$.
As indicated in Eq.~\eqref{eq:sojourn_probability}, the sojourn probability
 depends on the initial distribution $\Xvec_{\tinitial}\sim\Pinitial$ inside the tube;
 since we are mostly interested in the temporal decay rate of the sojourn probability,
 which for small tube radius is only affected by the initial condition for a short initial relaxation time
 $\tau_{\mathrm{rel}} \sim R(0)^2$, we suppress the dependence on initial conditions in the following.
 
The decay of the sojourn probability is described by the instantaneous
rate at which stochastic trajectories leave the tubular neighborhood of $\trajvec$ for the first time,
$\aexit(t)$, as 
\begin{align}
\Ptube(t)&=
\label{eq:exit_rate_definition}
\exp\left[-\int_{0}^{t}\mathrm{d}s~\aexit(s)\right].
\end{align}
The sojourn probability is a functional of both the reference path $\trajvec(t)$
and the function $R(t)$ which specifies the time-dependence of the radius,
and is equivalently described by the functional
\begin{equation}
\label{eq:Sdef_intro}
	\Stub[\trajvec,R] \equiv \Stub_R[\trajvec] \equiv \Stub_R^{\trajvec} \equiv \int_{\tinitial}^{t}\mathrm{d}s~\aexit(s),
\end{equation}
which we refer to as tubular stochastic action
 because it describes experimentally observable sojourn probabilities.
The exit rate can similarly be interpreted as a finite-radius (tubular) 
stochastic action Lagrangian $\mathcal{L}^{\trajvec}_R \equiv \aexit$.
As emphasized in Eq.~\eqref{eq:Sdef_intro}, we indicate functional dependences interchangeably
by square brackets and via sub- or superscripts.

In the following we focus on the case where the norm used to define
the tube is the standard Euclidean norm,
$||\xvec ||_2 \equiv \sqrt{x_1^2 + x_2^2 + \cdots + x_N^2}$,
and where the reference path $\trajvec$ is twice continuously differentiable.
To consistently speak of power-series expansions in the time-dependent radius $R(t)$,
we furthermore assume that $R(t) = R_0 r(t)$ for a reference radius $R_0$ and a dimensionless
differentiable function $r(t)$.
We then interpret power series in $R(t)$ as power series in $R_0$, 
meaning that we scale the radius uniformly (independent of time) along the path.

As we show in more detail  in Sect.~\ref{sec:exit_rate_in_terms_of_FP_spectrum}
below 
and 
in
App.~\ref{app:N_dim_theory}, 
the exit rate can be expanded as
a perturbation series in $R_0$ 
for small tube radius, yielding
\begin{align}
\mathcal{L}^{\trajvec}_R(t) \equiv \aexit(t) & =
\frac{\alpha_{\mathrm{free}}^{\trajvec}(t)}{R(t)^2} +
\alpha^{\trajvec,(0)}(t)
\label{eq:axit_result_intro} 
+\alpha^{\trajvec,(2)}R^2(t)
\\ & \qquad 
+\mathcal{O}(R_0^{4}),
\nonumber 
\end{align}
where
\begin{equation} 
\label{eq:afree_intro}
\frac{\alpha_{\mathrm{free}}^{\trajvec}(t) }{R(t)^2}
\equiv 
\frac{f(\Dmat (\trajvec(t),t))}{R(t)^{2}}.
\end{equation}
At time $t$, Eq.~\eqref{eq:afree_intro} is the steady-state free-diffusion exit rate 
from an $N$-dimensional ball of radius $R(t)$
for It\^{o}-Langevin dynamics Eq.~\eqref{eq:Ito_process_definition} 
with 
vanishing drift and
a spatially constant diffusion tensor
 $\Dmat (\trajvec(t),t)$.
The function $f(\underline{\bm{M}})$ 
is defined 
for a symmetric full-rank matrix $\underline{\bm{M}}$ with components $M_{ij}$
as the smallest negative
 eigenvalue of the anisotropic Laplace operator $M_{ij}\nabla_i \nabla_j$, with 
  domain
 the unit ball and absorbing boundary conditions,
 see App.~\ref{app:N_dim_spectrum} for more details.
While at time $t$ the free-diffusion exit rate Eq.~\eqref{eq:afree_intro} scales as $1/R(t)^2$
and only depends on $\Dmat(\trajvec(t),t)$,
the term
$\alpha^{\trajvec,(0)}(t)$ 
in Eq.~\eqref{eq:axit_result_intro} is of order $R(t)^0$
and
depends on $\avec(\trajvec(t),t)$, $\Dmat (\trajvec(t),t)$,
as well as their spatial and temporal derivatives up to second order,
 evaluated at 
$(\trajvec(t),t)$.
Note that, as we will see in our explicit one-dimensional example further below,
both $\alpha^{\trajvec,(0)}$ and $\alpha^{\trajvec,(2)}$ 
can depend on $\dot{r}/r$.

The It\^{o}-Langevin Eq.~\eqref{eq:Ito_process_definition} is 
on short length- and time scales
dominated by the random noise term,
which explains that  
for small radius the exit rate Eq.~\eqref{eq:axit_result_intro} is dominated by
 the instantaneous steady-state free-diffusion  exit rate Eq.~\eqref{eq:afree_intro}
 \cite{kappler_stochastic_2020}.
This rate diverges as the radius approaches zero, 
which via Eq.~\eqref{eq:exit_rate_definition} implies that the probability 
of any individual path vanishes.
On the other hand,
 the probability Eq.~\eqref{eq:exit_rate_definition} 
at finite radius is 
a physical observable,
as was noted before in the context of additive noise \cite{kappler_stochastic_2020}.
The expansion Eq.~\eqref{eq:axit_result_intro} 
allows to calculate this tube probability for small-but-finite time-dependent radius $R(t)$, 
and hence to
quantify pathway probabilities for diffusive trajectories in an experimentally measurable way.

\subsection{Exit rate in terms of FP spectrum}
\label{sec:exit_rate_in_terms_of_FP_spectrum}

To derive the perturbation series Eq.~\eqref{eq:axit_result_intro} for the exit rate
for a given reference path $\trajvec(t)$ and tube radius $R(t)$,
we
consider 
the equivalent description of the stochastic process Eq.~\eqref{eq:Ito_process_definition}
 inside
the tubular neighborhood 
of the reference path
via the
FP Eq.~\eqref{eq:ND_FP_eq_intro}.
At time $t$, the spatial domain for Eq.~\eqref{eq:ND_FP_eq_intro} is then
\begin{align}
\xvec\in\BR(t)&
\label{eq:tubular_neighborhood_at_time_t}
\equiv\left\{ \,\xvec\,\,\setbar\,\, || \xvec - \trajvec(t)||_2<R(t)\,\right\}, 
\end{align}
which is indicated in Fig.~\ref{fig:tubular_ensemble} as a gray shaded area.
The solution $\Ptube(\xvec,t)$ to the FPE is subject to absorbing boundary conditions at the tube boundary,
$\Ptube(\xvec,t)=0$ for all $\vecx\in\partial\BR(t)$.
 $\Ptube(\xvec,t)$ then
describes the spatial distribution of all stochastic trajectories $\Xvec_t$ that have never left
the tubular neighborhood until time $t$. 
Note that in Eq.~\eqref{eq:tubular_neighborhood_at_time_t} we consider the standard Euclidean norm,
so that $\BR$ describes a moving ball in $\mathbb{R}^N$, 
with instantaneous radius $R(t)$ and center $\trajvec(t)$.

To obtain Eq.~\eqref{eq:axit_result_intro}, 
 we use the
 same strategy as in a recent derivation  of finite-radius tubular exit rates for
 Langevin dynamics with isotropic additive diffusivity $\Dmat = D_0\underline{\mathbb{1}}$, where $D_0$ is a positive scalar
  and $\underline{\mathbb{1}}$ is the unit matrix \cite{kappler_stochastic_2020}.
We here shortly summarize the derivation, and refer the reader to App.~\ref{app:N_dim_theory} 
and Ref.~\cite{kappler_stochastic_2020} for more details.
Our derivation assumes that $\trajvec$ is twice continuously differentiable,
and that $R(t) = R_0 r(t)$ so that $\dot{R}/R = \dot{r}/r$ is independent of $R_0$.
To derive Eq.~\eqref{eq:axit_result_intro}
we first introduce
dimensionless streaming coordinates that move along the tube center;
this removes the time-dependence of the boundary conditions.
In the streaming coordinate system we then
 project the FPE onto the instantaneous eigenbasis of the FP operator.
In this eigenbasis, we derive an approximate solution of the FPE as a perturbation
series in the small tube radius, 
using an approach similar to time-dependent perturbation theory in quantum mechanics \cite{ballentine_quantum_2010}.
This perturbative solution $\Ptube(\xvec,t)$ is, 
after an initial relaxation timescale $\tau_{\mathrm{rel}} \sim R(\tinitial)^2$, 
dominated by the decay of the slowest-decaying eigenfunction.
For our perturbative solution we assume that the diffusivity tensor has full rank, which implies that
 all degrees of freedom are for small length- and time-scales dominated by random forces.

The sojourn
probability up to time $t$ is simply the survival probability,
which we calculate as the spatial integral over the solution $\Ptube(\xvec,t)$ of the
 FPE as
$\Ptube(t)
=\int_{\BR(t)}\mathrm{d}^{N}\xvec\,\Ptube(\xvec,t)$.
From the survival probability 
we finally evaluate
the instantaneous exit rate, defined in Eq.~\eqref{eq:exit_rate_definition}, 
as
\begin{equation}
\aexit(t)=-\frac{\Ptubedot(t)}{\Ptube(t)}.\label{eq:inst_exit_rate}
\end{equation}
Upon evaluating this expression using the perturbative
 solution of the FPE, 
the exit rate Eq.~\eqref{eq:axit_result_intro} follows.
In App.~\ref{app:N_dim_theory} we give resulting expressions for
$f$, $\alpha^{\trajvec,(0)}$, $\alpha^{\trajvec,(2)}$,
in terms of the instantaneous FP spectrum inside the tube.
In particular we show that the equation that determines $f$ at time $t$
only depends on $\Dmat(\trajvec(t),t)$ and neither on spatial derivatives of the
diffusivity tensor nor on the drift,
which justifies the notation $f(\Dmat (\trajvec(t),t))$ in Eq.~\eqref{eq:afree_intro}.

\subsection{Asymptotic ratios of tube probabilities}
\label{sec:asymptotic_tube_probabilities_N}

While the probability to observe any individual path is zero,
ratios of probabilities for individual paths can be defined 
in special cases.

One such case are systems
 with a constant isotropic diffusion tensor
$\Dmat \equiv D_0\, \underline{\mathbb{1}} \equiv const.$,
where $D_0$ is positive and $\underline{\mathbb{1}}$ denotes the unit matrix,
and considering tubes of time-independent radius, $R(t) \equiv R_0 = const.$,
with respect to the standard Euclidean 
norm \cite{stratonovich_probability_1971,
durr_onsager-machlup_1978,
fujita_onsager-machlup_1982,ikeda_stochastic_1989,
ito_probabilistic_1978,williams_probability_1981,
horsthemke_onsager-machlup_1975,
kappler_stochastic_2020}.
In this scenario the
function $f$ in Eq.~\eqref{eq:afree_intro} is
independent of the path $\trajvec$,
so that the
 subleading-order term $\LagrangianOnsagerMachlup \equiv \alpha^{\trajvec,(0)}$
quantifies relative path probabilities \cite{kappler_stochastic_2020}.
The ratio of path probabilities for 
 two paths $\trajvec$, $\trajTwovec$
 is then defined as vanishing-radius limiting ratio of sojourn probabilities,
and quantified via a stochastic action $\SOM$ 
as \cite{stratonovich_probability_1971,durr_onsager-machlup_1978,
ikeda_stochastic_1989,kappler_stochastic_2020,
horsthemke_onsager-machlup_1975,
ito_probabilistic_1978,williams_probability_1981,
fujita_onsager-machlup_1982}
\begin{align}
\frac{e^{-\SOM[\bm{\traj}]}}{e^{-\SOM[\bm{\trajTwo}]}}
\label{eq:limiting_ratio_of_tube_probabilities}
&\equiv\lim_{R_0\rightarrow0}\frac{\Ptube(\tfinal)}{\PtubeTwo(\tfinal)},
\end{align}
where the Onsager-Machlup (OM) action $\SOM[\trajvec]$, which is a functional
of the twice continuously differentiable path $\trajvec$, is found to be 
\begin{align}
\SOM[\bm{\trajvec}]&=
\int_{\tinitial}^{\tfinal}\mathrm{d}t~\alpha^{\trajvec,(0)}(t)
\equiv
\int_{\tinitial}^{\tfinal}\mathrm{d}t~\LagrangianOnsagerMachlup(\bm{\trajvec}(t),\dot{\bm{\trajvec}}(t),t),\label{eq:OnsagerMachlupAction}
\end{align}
with the OM Lagrangian
\begin{align}
\LagrangianOnsagerMachlup(\vectraj,\dot{\bm{\traj}})\equiv \alpha^{\trajvec,(0)} 
&= 
\frac{1}{4D_0}\left[\dot{\bm{\traj}}-\avec(\trajvec)\right]^{2}+\frac{1}{2}\grad\cdot\avec(\vectraj).\label{eq:OnsagerMachlupLagrangian}
\end{align}
For additive isotropic noise, ratios of path probabilities can thus be defined as limits 
of temporally-constant-radius sojourn probabilities, and the OM Lagrangian quantifies such 
ratios \cite{stratonovich_probability_1971,
ito_probabilistic_1978,williams_probability_1981,
horsthemke_onsager-machlup_1975,
fujita_onsager-machlup_1982,kappler_stochastic_2020,
durr_onsager-machlup_1978}.
In view of Eqs.~\eqref{eq:Sdef_intro}, \eqref{eq:axit_result_intro},
the OM action $\SOM$ and Lagrangian $\LagrangianOnsagerMachlup$
are the order-$R^0$ contributions of the finite-radius tubular action
$\Stub$ and its associated 
tubular Lagrangian for
the special case of 
additive isotropic noise
and
a
 temporally constant tube radius.

For a state-dependent diffusion tensor, however,
the limit in Eq.~\eqref{eq:limiting_ratio_of_tube_probabilities} in general does not yield meaningful results.
In that case, it follows from substituting Eqs.~\eqref{eq:axit_result_intro}, \eqref{eq:afree_intro},
into Eq.~\eqref{eq:exit_rate_definition} that
\begin{align}
\ln\frac{\Ptube(\tfinal)}{\PtubeTwo(\tfinal)}
\nonumber=&
-\int_{\tinitial}^{\tfinal}\mathrm{d}s\,\left[ 
\frac{f(\Dmat(\trajvec(s),s))}{R_{\trajvec}^{2}(s)} - \frac{f(\Dmat(\trajTwovec(s),s))}{R_{\trajTwovec}^{2}(s)} 
\right]
\\ & 
\label{eq:ratio_of_path_probabilities}
-\int_{\tinitial}^{\tfinal}\mathrm{d}s\,\left[ 
\alpha^{\trajvec,(0)}(s) - \alpha^{\trajTwovec,(0)}(s)
\right]
\\ &
 \nonumber
+ \mathcal{O}\left(R_0^2\right)
\end{align}
where 
we allow for 
different choices of $R(t)$ along 
the paths, 
which we indicate by the notation
 $R_{\trajvec}(t) = R_0 r_{\trajvec}(t)$, 
$R_{\trajTwovec}(t)= R_0 r_{\trajTwovec}(t)$.
Since $f$ is a function of the local diffusivity along the path, 
 it follows from
 Eq.~\eqref{eq:ratio_of_path_probabilities}
 that in general
 the difference of the free-diffusion exit rates does not vanish,
so that in the limit
$R_0 \rightarrow 0$
 the expression Eq.~\eqref{eq:ratio_of_path_probabilities} 
diverges as $1/R_0^2$.
The physical origin of this divergence is that
in a region with low diffusivity a particle is less likely to diffuse away from a reference
path, as compared to a region with large diffusivity \cite{kappler_stochastic_2020}.

While in general the log-ratio Eq.~\eqref{eq:ratio_of_path_probabilities} diverges in
 the limit $R_0 \rightarrow 0$,
we can obtain a finite limit by carefully choosing the path-dependent tube radius
$R_{\trajvec}$, $R_{\trajTwovec}$.
To demonstrate this
we consider the tube radius
\begin{equation}
\label{eq:tube_radius_scaling_away_free_diffusion}
R_{\trajvec}(t) = R_0 \sqrt{ \frac{f(\Dmat(\trajvec(t),t)}{f_0}},
\end{equation}
 with constants $R_0$, $f_0$,
 and also the analogous path-dependent radius for $R_{\trajTwovec}$
 with the same constants $R_0$, $f_0$. 
From Eq.~\eqref{eq:ratio_of_path_probabilities} it is then evident that 
the two leading order free-diffusion terms become path-independent, and hence
their difference cancels. 
The limit of vanishing 
tube radius $R_0 \rightarrow 0$ is then finite,
\begin{align}
\lim_{R_0 \rightarrow 0}\ln\frac{\Ptube(\tfinal)}{\PtubeTwo(\tfinal)}
=&
\label{eq:ratio_of_path_probabilities_scenario2}
-\int_{\tinitial}^{\tfinal}\mathrm{d}s\,\left[ 
\alpha^{\trajvec,(0)}(s) - \alpha^{\trajTwovec,(0)}(s)
\right],
\end{align}
which can be rewritten in a form more reminiscent of 
Eq.~\eqref{eq:limiting_ratio_of_tube_probabilities}
 as
\begin{align}
\frac{e^{-\SR[\bm{\traj}]}}{e^{-\SR[\bm{\trajTwo}]}}
\label{eq:limiting_ratio_of_tube_probabilities_scaled}
&\equiv\lim_{R_0\rightarrow0}\frac{\Ptube(\tfinal)}{\PtubeTwo(\tfinal)},
\end{align}
with the action
\begin{align}
\label{eq:SR_action}
\SR[\trajvec] &= \int_{\tinitial}^{\tfinal}\mathrm{d}s~\alpha^{\trajvec,(0)}(s),
\end{align}
where we choose the subscript r because this action is based on the local
\textbf{r}escaling Eq.~\eqref{eq:tube_radius_scaling_away_free_diffusion}.
If the  diffusion tensor is constant and isotropic, then 
according to Eq.~\eqref{eq:tube_radius_scaling_away_free_diffusion}
the radius is constant and $\SR \equiv \SOM$. 
Despite the formal similarity between
Eqs.~\eqref{eq:limiting_ratio_of_tube_probabilities} and
\eqref{eq:limiting_ratio_of_tube_probabilities_scaled},
there is an important difference between the two limits.
Equation \eqref{eq:limiting_ratio_of_tube_probabilities}
considers the 
exit rate from a constant-radius tube with respect to the ambient Euclidean metric,
which is a natural choice when observing experimental data. 
On the other hand, in Eq.~\eqref{eq:limiting_ratio_of_tube_probabilities_scaled}
we consider the path-dependent tube radius 
Eq.~\eqref{eq:tube_radius_scaling_away_free_diffusion},
which is designed to scale away the leading-order
differences in the small-radius exit rate.
To measure the exit rate from a tube with radius
Eq.~\eqref{eq:tube_radius_scaling_away_free_diffusion}
in an experiment,
both the diffusion tensor $\Dmat$ 
 along the path $\trajvec$ 
and the explicit functional form of $f$ 
need to be known
to
 evaluate $R_{\trajvec}$.
The ratios of path probabilities Eq.~\eqref{eq:limiting_ratio_of_tube_probabilities_scaled}
are thus
not straightforwardly related to what one would measure in an experiment.
Furthermore, the choice
Eq.~\eqref{eq:tube_radius_scaling_away_free_diffusion} 
is not
the only construction that leads to a finite limiting-ratio of tube probabilities.
Another possibility is to
define a tube via the metric induced by the diffusion tensor
$\Dmat$ \cite{ventsel_small_1970}
which corresponds to considering a moving
 ellipsoid in $\mathbb{R}^N$, whose principal axes vary along the reference path
 in such a way that the steady-state free-diffusion exit rate is independent of 
 the chosen path.
 Only for one-dimensional systems, $N=1$, where ellipsoids and 
 balls are identical and simply given by intervals,
 do these two constructions
 lead to the same tube.
Yet another construction for obtaining a finite limiting ratio is 
due to Stratonovich \cite{stratonovich_probability_1971},
and leads to the standard Lagrangian for multiplicative noise.
As we show explicitly    in Sect.~\ref{sec:results_for_one_dimensional_system}
for a one-dimensional system,
the underlying geometrical idea of the Stratonovich construction
 is to perform a nonlinear coordinate 
transformation such that the diffusivity becomes constant,
and then to consider a constant-radius tube in this coordinate system.
While all three methods (Eq.~\eqref{eq:tube_radius_scaling_away_free_diffusion}, 
using the metric induced by the
diffusion tensor, the Stratonovich construction) lead to finite limiting
ratios of tube probabilities, they in general correspond to different tubes in spacetime;
therefore they lead to different finite-radius sojourn probabilities, and consequently
 different actions.
\textcolor{black}{
All three methods also require knowledge of the diffusivity profile, which might not be 
readily available when working with measured data.}

The technical difficulties and  ambiguities in
extending the vanishing-radius limit of tube probabilities
Eq.~\eqref{eq:limiting_ratio_of_tube_probabilities}
to systems with multiplicative noise,
together with the fact that any individual path has vanishing probability,
suggests that instead of considering the limit of vanishing tube radius,
 focus should be put on the finite-radius sojourn probability Eq.~\eqref{eq:exit_rate_definition}
which describes observable events of positive probability.

We note that while in general the log-ratio Eq.~\eqref{eq:ratio_of_path_probabilities}
diverges in the limit of vanishing tube radius,
an important exception 
is the case where $\trajTwovec$, $R_{\trajTwovec}$ are the time reverse of $\trajvec$, $R_{\trajvec} \equiv R$, i.e.~where
$\trajTwovec(t) \equiv \overset{\leftarrow}{\trajvec} (t) \equiv \trajvec( \tfinal - t)$,
$R_{\trajTwovec}(t) \equiv R( \tfinal - t)$,
and where we assume that for the reverse path
also all explicit time-dependences in $\avec$, $\Dmat$ are reversed,
as is customary when considering irreversibility in stochastic thermodynamics \cite{seifert_entropy_2005,seifert_stochastic_2012}.
The leading order terms in Eq.~\eqref{eq:ratio_of_path_probabilities}
then cancel and we obtain
\begin{align}
\label{eq:ratio_of_path_probabilities_reverse}
\lim_{R_0\rightarrow 0} 
\ln\frac{\Ptube(\tfinal)}{P_{R}^{\overset{\leftarrow}{\trajvec}}(\tfinal)}
&=
-\int_{\tinitial}^{\tfinal}\mathrm{d}s\,\left[ 
\alpha^{\trajvec,(0)}(s) - \alpha^{\overset{\leftarrow}{\trajvec},(0)}(s)
\right].
\end{align}
The limiting ratio of sojourn probabilities for a pair of forward and reverse path
 is thus generally finite,
and can hence be used to quantify pathwise irreversibility.
Further below we evaluate the limit Eq.~\eqref{eq:ratio_of_path_probabilities_reverse}
explicitly for one-dimensional systems, and relate the result to the usual
pathwise definition of the entropy production \cite{seifert_entropy_2005,
bo_functionals_2019,
kappler_measurement_2022,
cates_stochastic_2022}.

\subsection{Most probable tubes and most probable paths}

One application of path probabilities is determining most probable paths \cite{durr_onsager-machlup_1978,adib_stochastic_2008},
which especially in the case of low noise can provide information about the typical
behavior of the stochastic dynamics \cite{ventsel_small_1970}.
For a finite radius we consider the most probable tube (MPT) center $\trajvec^*$ connecting an initial position $\xvec_0$
at time $t = 0$
and a final position $\xvec_f$ at time $t= \tfinal$.
The MPT center $\trajvec^*$ is the path that maximizes 
the sojourn probability, and 
is obtained by minimizing the action Eq.~\eqref{eq:Sdef_intro},
\begin{align}
\label{eq:def_most_probable_tube}
	\trajvec^* \equiv  \underset{\trajvec}{\mathrm{argmin}} \,\Stub[\trajvec,R_{\trajvec}],
\end{align}
where the minimization is over all twice differentiable paths which
fulfill $\trajvec(0) = \xvec_0$, $\trajvec(\tfinal) = \xvec_f$,
and where 
the time-dependent tube radius $R_{\trajvec}$ may depend on the path (of course the details of this path-dependence
need to be specified before the functional Eq.~\eqref{eq:def_most_probable_tube} can be minimized).
After $\trajvec^*$ has been obtained, the finite sojourn probability to observe any trajectory that remains
within the tube is calculated via Eq.~\eqref{eq:exit_rate_definition}.
A most probable path (MPP) may be defined 
from Eq.~\eqref{eq:def_most_probable_tube}
as most MPT center in the limit 
of vanishing radius; 
as we see explicitly in Sect.~\ref{sec:most_probable_tube} below,
the result of course depends on the exact form of $R_{\trajvec}(t)$.

For additive noise
 with a constant isotropic diffusion tensor
$\Dmat \equiv D_0\, \underline{\mathbb{1}} \equiv const.$,
and a constant radius, $R(t) \equiv R_0 = const.$,
 the limit $R_0 \rightarrow 0$ in Eq.~\eqref{eq:def_most_probable_tube}
is equivalent to finding a path $\trajvec^*$ that minimizes
 the OM action Eq.~\eqref{eq:OnsagerMachlupAction}.
For the general Langevin Eq.~\eqref{eq:Ito_process_definition} with multiplicative noise,
we see from  Eqs.~\eqref{eq:axit_result_intro}, \eqref{eq:afree_intro}, 
 that 
in the limit of vanishing time-independent tube radius, the MPP in general 
 minimizes the average free-diffusion exit rate along the path.
Thus, because diffusive dynamics is on short length- and time scales dominated by the 
random noise term in Eq.~\eqref{eq:Ito_process_definition}, 
for an It\^{o}-Langevin dynamics with multiplicative noise the drift term is completely
 irrelevant for the MPP \cite{kappler_stochastic_2020}.
For a one-dimensional system this is discussed further below and in the
accompanying Ref.~\cite{accompanying_paper_short}.

\section{One-dimensional systems}
\label{sec:results_for_one_dimensional_system}

\subsection{Exit rate}
\label{sec:1D_exit_rate}

We now consider a one-dimensional system, $N=1$,
for which the It\^{o} Eq.~\eqref{eq:Ito_process_definition}
becomes
\begin{align}
\mathrm{d}{X}_{t}&= a(X_t,t)
\,\mathrm{d}t
\label{eq:Langevin_1D}
+\sqrt{2D(X_t,t)} \,
				 \mathrm{d}W_{t},
\end{align}
where we use that for one-dimensional systems the noise strength 
is expressed in terms of the diffusivity as $b(x,t)= \sqrt{2D(x,t)}$.
Similarly, the FP Eq.~\eqref{eq:ND_FP_eq_intro} becomes
\begin{align}
\partial_{t}P&=
-\partial_x\left(
a 
P\right)
\label{eq:1D_FP_eq} 
+\partial_x^2 \left(D P\right).
\end{align}

\begin{figure*}[ht]
\centering \includegraphics[width=1\textwidth]{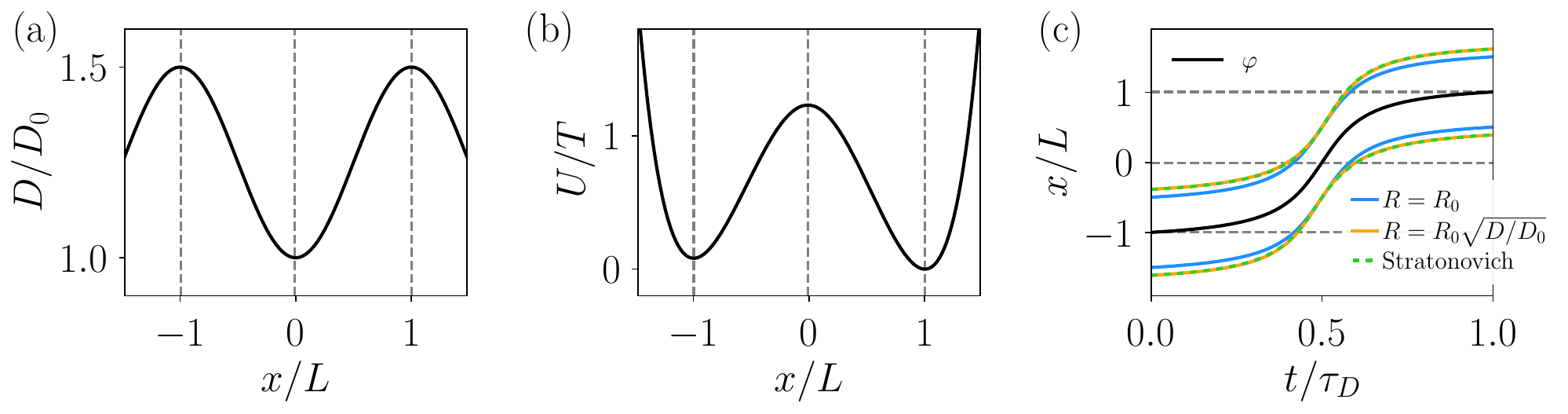} 
\caption{ \label{fig:potential_and_trajectory} 
Diffusivity profile, potential, and reference path
used for the one-dimensional examples in Sect.~\ref{sec:results_for_one_dimensional_system}.
Subplot (a) depicts the double well potential
Eq.~\eqref{eq:boltzmann_inversion} obtained
using Eqs.~\eqref{eq:P_ss}, \eqref{eq:D_example}, 
\eqref{eq:drift_potential}.
Subplot (b) shows the diffusivity profile Eq.~\eqref{eq:D_example}.
Subplot (c) depicts the reference path $\traj$ defined in
Eq.~\eqref{eq:reference_path}.
Around the
reference path, 
 the boundaries of the respective tubes for the three scenarios 
considered in Sect.~\ref{sec:1D_exit_rate} are plotted.
For better visibility, all tube radii are increased by a factor of 5 for the plot.
In all subplots, gray dashed lines denote two potential barrier minima $x = \pm L$.
}
\end{figure*}

For one-dimensional systems, we calculate the FP spectrum  explicitly
in terms of the drift and diffusivity, as discussed in more detail  
in App.~\ref{app:one_dim}.
We use the resulting perturbative spectrum 
to evaluate the exit rate Eq.~\eqref{eq:axit_result_intro} to order $R_0^2$.
The two lowest-order terms in this expansion are
\begin{align}
\label{eq:afree_1D}
\frac{
\aexitfree^{\traj} (t)
}{
R(t)^2
}&=   \frac{\pi^2}{4}\frac{D(\traj(t),t)}{R(t)^2}
\end{align}
and 
\begin{widetext}
\begin{align}
\mathcal{L}^{\traj,(0)} &\equiv
\alpha^{\traj,(0)} = 
\label{eq:Lagrangian_1D}
 \frac{1}{4D} \left( \dot{\traj} - a + \partial_x D\right)^2
+\frac{1}{2} \partial_x a
- \frac{1}{16}\left( \pi^2 - 1 \right) \frac{ \left( \partial_x D\right)^2}{D}
+ \frac{1}{4} \left( \frac{ \pi^2}{6} - 1 \right) \partial_x^2 D
- \frac{1}{2} \frac{\dot{r}}{r},
\end{align}
\end{widetext}
where $D$, $a$ and their derivatives are evaluated at $(x,t) \equiv (\traj(t),t)$, 
the radius is in general time-dependent, $R(t) \equiv R_0r(t)$, where $r(t)$ may depend on $\traj$.
The explicit expression for the quadratic term $\alpha^{\traj,(2)}$
in Eq.~\eqref{eq:axit_result_intro} is lengthy, and we provide it
in App.~\ref{app:one_dim}.
We furthermore provide a python package which includes
all analytical results derived in this paper as symbolic expressions \cite{module_pytubular}.

Our result Eq.~\eqref{eq:Lagrangian_1D} is different from 
the Lagrangians derived
for multiplicative noise in the 
literature \cite{stratonovich_probability_1971,
durr_onsager-machlup_1978,
cugliandolo_building_2019,stratonovich_probability_1971,horsthemke_onsager-machlup_1975,
fujita_onsager-machlup_1982,ikeda_stochastic_1989,
ito_probabilistic_1978,williams_probability_1981}.
The relevance of our result $\mathcal{L}^{\traj,(0)}$ is that it appears
as a term
in the perturbative expansion of the exit rate Eq.~\eqref{eq:axit_result_intro}, 
and hence is a physical observable.

\subsection{Limiting probability ratio for forward-backward path pair}
\label{sec:limiting_ratio}

As we discussed in Sect.~\ref{sec:asymptotic_tube_probabilities_N},
for two paths $\traj$, $\trajTwo$
the log-ratio of tube probabilities Eq.~\eqref{eq:ratio_of_path_probabilities}
 in general diverges in the limit of vanishing tube radius.
However, if for $\trajTwo$ we consider the time reverse of $\traj$,
 i.e.~$\trajTwo(t) \equiv \overset{\leftarrow}{\traj}(t) = \traj(\tfinal -t)$,
 and also reverse all explicit time-dependences in $D$, $a$, $R$,
 then from Eqs.~\eqref{eq:ratio_of_path_probabilities_reverse},
\eqref{eq:Lagrangian_1D},
we obtain
 \begin{align}
\label{eq:1D_limit_radius_zero_reverse}
\lim_{R_0 \rightarrow 0}
\ln \frac{P_R^{\traj}(\tfinal)}{P_{R}^{\overset{\leftarrow}{\traj}}(\tfinal)} &=
\int_{\tinitial}^{\tfinal}\mathrm{d}t\,\left. \frac{
 a - \partial_x D}{D}\right|_{(\traj(t),t)} \dot{\traj}(t)
+  \ln \frac{r(\tfinal) }{r(\tinitial)}.
\end{align}

To cast the right-hand side of Eq.~\eqref{eq:1D_limit_radius_zero_reverse}
in a more familiar form, we note that
the instantaneous steady-state solution of Eq.~\eqref{eq:1D_FP_eq} subject to 
instantaneous no-flux boundary conditions at the system boundary (i.e.~the full
spatial domain of the system without assuming a tube) is given by \cite{gardiner_stochastic_2009}
\begin{align}
\label{eq:P_ss}
P_{\mathrm{ss}}(x,t) &= \frac{\mathcal{N}(x_0)}{D(x,t)}\exp\left[ 
\int_{x_0}^{x}dx'\,\frac{a(x',t)}{D(x',t)},
\right]
\end{align}
where $x_0$ is an arbitrary point in the domain and $\mathcal{N}(x_0)$ is 
a normalization constant which ensures that $P_{\mathrm{ss}}$ is a proper probability density.
We define the instantaneous potential $U(x,t)$ via
 Boltzmann inversion of the steady-state solution,
\begin{equation}
\label{eq:boltzmann_inversion}
U(x,t) = -T\ln[P_{\mathrm{ss}}(x,t)L],
\end{equation}
 where $T$ is the absolute temperature in units of energy
 and $L$ is an arbitrary length scale to render the argument of the
 logarithm function dimensionless.
 
 With these definitions, 
 we rewrite the limit Eq.~\eqref{eq:1D_limit_radius_zero_reverse}
 as
 \begin{align}
\lim_{R_0 \rightarrow 0}
\ln \frac{P_R^{\traj}(\tfinal)}{P_{R}^{\overset{\leftarrow}{\traj}}(\tfinal)} &=
- \frac{\Delta U}{T} 
+ \frac{1}{T} \int_0^{t_f}dt\,\left[ (\partial_t U) (\traj(t),t) \right]
\nonumber
\\ & \qquad 
+  \ln \frac{r(\tfinal) }{r(\tinitial)},
\label{eq:1D_limit_radius_zero_reverse_2}
\end{align}
where $\Delta U = U(\traj(\tfinal),\tfinal) - U(\traj(\tinitial),\tinitial)$.
 
If the drift and diffusivity are independent of time, then $\partial_t U =0$ so that 
the second term on the right-hand side of
Eq.~\eqref{eq:1D_limit_radius_zero_reverse_2} vanishes. 
If furthermore the tube radius at the initial and final time are identical, $r(\tfinal) = r(\tinitial)$,
which includes the scenario of a constant tube radius $R(t) \equiv R_0 = const.$,
then also the third term on the right-hand side 
of Eq.~\eqref{eq:1D_limit_radius_zero_reverse_2} is zero.
In that case, 
the remaining potential difference on the right-hand side of Eq.~\eqref{eq:1D_limit_radius_zero_reverse_2}
represents the familiar formula for the entropy production along
the path $\traj$, which for multiplicative noise  
has previously
 been derived via time-discretization path-integral methods \cite{cates_stochastic_2022}.

\begin{figure*}[ht]
\centering \includegraphics[width=\textwidth]{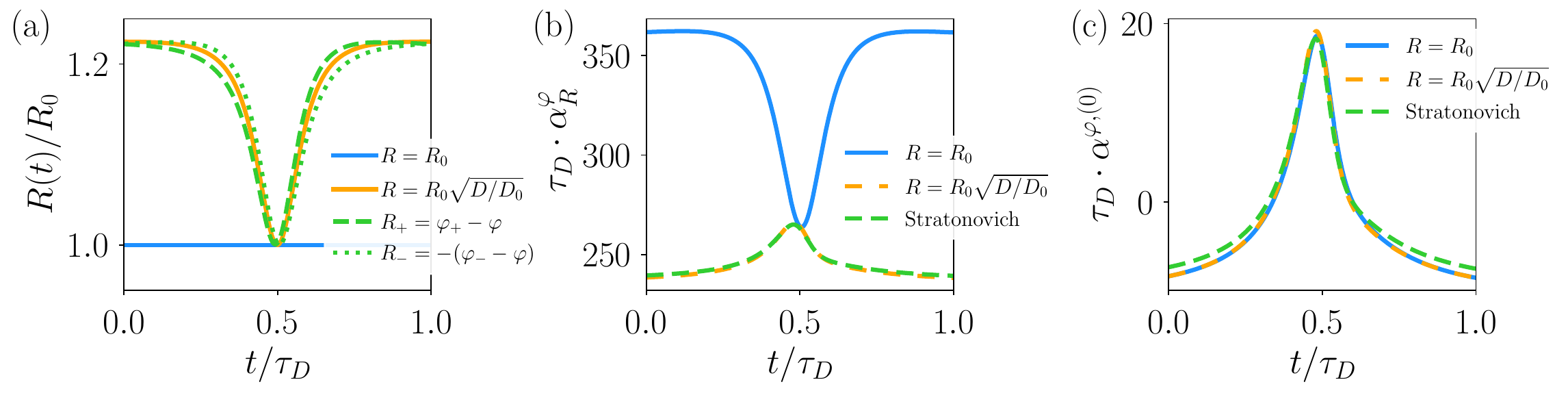} 
\caption{ \label{fig:radius_and_theoretical_rates} 
Tube radius and exit rates for the one-dimensional 
example system considered in Sect.~\ref{sec:results_for_one_dimensional_system}.
Subplot (a) shows the time-dependent tube radius for constant tube radius (scenario 1; blue horizontal solid line),
constant free-diffusion exit rate (scenario 2; orange solid line),
and the distance from the reference path to the two interval bounds for the Stratonovich construction
(scenario 3; green dashed and dotted lines). For all scenarios, we use $R_0/L = 0.1$.
Subplot (b)
shows the perturbative exit rate Eq.~\eqref{eq:axit_result_intro} to order $R^0$ (inclusive) for
 constant tube radius (scenario 1; blue solid line),
constant free-diffusion exit rate (scenario 2; orange solid line),
and the Stratonovich construction
(scenario 3; green dotted line). 
Subplot (c) shows the  first correction to the steady-state free-diffusion exit rate,
 $\mathcal{L}^{\traj,(0)} \equiv \alpha^{\traj,(0)}$,
for the three scenarios considered in subplot (b);
these corrections are given by   Eq.~\eqref{eq:scenario1_Lagrangian} (scenario 1, blue solid line),
 Eq.~\eqref{eq:scenario2_Lagrangian} (scenario 2, orange dashed line),
and  Eq.~\eqref{eq:scenario3_Lagrangian} (scenario 3, green dotted line).
}
\end{figure*}

\subsection{Numerical example: Particle in an asymmetric double well}
\label{sec:numerical_example}

For our numerical example we fix a length scale $L$ and a
diffusivity scale $D_0$, which defines the diffusive
time scale $\td  = L^2/D_0$.
We consider a diffusivity profile
\begin{equation}
\label{eq:D_example}  
	D(x) =  \frac{D_0}{4} \left[ 5 - \cos\left(\pi \frac{x}{L}\right)\right],
\end{equation}
which features a locally maximal diffusivity $D(x=\pm L) = 3D_0/2$ at $x = \pm L$
and a locally minimal diffusivity $D(x=0) = D_0$ at $x=0$. 
This diffusivity profile 
 is shown in Fig.~\ref{fig:potential_and_trajectory} (a).
 
We furthermore consider a gradient drift profile 
$a(x) = - (\partial_x U_a)(x)$,
with corresponding potential
\begin{align}
\label{eq:drift_potential}
U_a(x) &= a_0 \left[ \left(\frac{x}{L}\right)^2 - 1 \right]^2 - a_1 \frac{x}{L},
\end{align}
where for our numerical results we use $a_0 = 2 L^2/T$ and $a_1 = a_0/20$.
Since $a_1 \ll a_0$, Eq.~\eqref{eq:drift_potential} 
describes a slightly asymmetric double well.
We include this asymmetry to avoid any potential issues of 
degeneracy in determining barrier-crossing MPT centers, 
as discussed further below.

Note that because the noise is multiplicative, 
the drift potential Eq.~\eqref{eq:drift_potential} is not 
proportional to
 the steady-state potential $U(x)$ defined in 
 Eq.~\eqref{eq:boltzmann_inversion}.
This is most clearly seen from Eq.~\eqref{eq:P_ss}, which relates the drift
$a(x) = - \partial_x U_a$ 
and the steady state $P_{\mathrm{ss}}(x)$.
We show the steady-state potential
Eq.~\eqref{eq:boltzmann_inversion}
 in Fig.~\ref{fig:potential_and_trajectory} (b),
where we observe that the two local minima of $U(x)$
are located at $\xl \approx - L$ and $\xr \approx L$.
 As Fig.~\ref{fig:potential_and_trajectory}  (a), (b) shows, 
the system we consider here is
qualitatively similar to the corrugated channel from
the accompanying Ref.~\cite{accompanying_paper_short}.

As reference path $\varphi$ we consider
\begin{align}
\traj(t)&=
\label{eq:reference_path}
\frac{\xr + \xl}{2}
\\ &\qquad \nonumber
+\frac{\xr - \xl}{2\arctan(\kappa/2)}\arctan\left[\kappa\cdot\left(\frac{t-t_f/2}{\tau_D}\right)\right],
\end{align}
which is a barrier crossing path 
that starts at time $t_i = 0$ at $\traj(t_i)=\xl$
and ends at time $t_f = \td$ at $\traj(t_f)=\xr$.
For the parameter $\kappa$, which sets the maximal speed at which
 $\traj$ crosses the barrier,
we use $\kappa=10$. 
We show the reference path Eq.~\eqref{eq:reference_path}
 in Fig.~\ref{fig:potential_and_trajectory} (c).

We now discuss the tubular exit rate for two particular choices of the time-dependent radius,
namely the scenarios of constant radius $R(t) \equiv R_0$, and
the scenario of constant free-diffusion exit rate Eq.~\eqref{eq:tube_radius_scaling_away_free_diffusion}.
Subsequently we relate the one-dimensional Stratonovich stochastic 
action Lagrangian \cite{stratonovich_probability_1971}
to tubular exit rates.

In this section we only consider the exit rate to order $R^0$ (inclusive).
In App.~\ref{app:one_dim} 
we compare the theoretical exit rates shown 
in this section to numerically evaluated exit rates, to demonstrate
 that 
for the present system and tube radius considered, 
the perturbative exit rate Eq.~\eqref{eq:axit_result_intro} to order $R^0$ 
approximates the actual exit rate well.

\textit{Scenario 1: Constant tube radius.}
For a tube with constant radius $R(t) \equiv R_0 = const.$, 
illustrated as horizontal blue solid line in Fig.~\ref{fig:radius_and_theoretical_rates} (a),
we obtain from Eqs.~\eqref{eq:afree_1D}, \eqref{eq:Lagrangian_1D}
that
\begin{align}
\label{eq:scenario1_free_diffusion}
	\frac{
		\aexitfree^{\traj}
	}{
		R^2
	}
	 &= \frac{\pi^2}{4}\frac{D}{R_0^2},
\\
\mathcal{L}_1^{\traj,(0)} &= 
 \frac{1}{4D} \left( \dot{\traj} - a + \partial_x D\right)^2
+\frac{1}{2} \partial_x a
\nonumber \\&  \qquad 
\label{eq:scenario1_Lagrangian}
- \frac{1}{16}\left( \pi^2 - 1 \right) \frac{ \left( \partial_x D\right)^2}{D}
+ \frac{1}{4} \left( \frac{ \pi^2}{6} - 1 \right) \partial_x^2 D,
\end{align}
where as before both $a$, $D$ and their spatial derivatives are evaluated 
along $\traj(t)$; the subscript 1 in Eq.~\eqref{eq:scenario1_Lagrangian} indicates 
that this is the Lagrangian for the first scenario we consider.
In Fig.~\ref{fig:radius_and_theoretical_rates} (b) we show the exit rate
Eq.~\eqref{eq:axit_result_intro} to order $R^0$,
evaluated along the path Eq.~\eqref{eq:reference_path}
 for a tube radius $R_0/L= 0.1$
 and using Eqs.~\eqref{eq:scenario1_free_diffusion},
 \eqref{eq:scenario1_Lagrangian}.
We observe that the exit rate 
takes on a minimum at time  $t \approx 0.5 \td$,
i.e.~when the path is close to the barrier top $x \approx 0$.
At this point also the diffusivity profile shown in Fig.~\ref{fig:potential_and_trajectory} (b)
displays a minimum;
this suggests that the radius $R_0/L = 0.1$ is so small that 
the exit rate Eq.~\eqref{eq:axit_result_intro} is already dominated by 
the
free-diffusion contribution Eq.~\eqref{eq:scenario1_free_diffusion},
which is proportional to $D$.
Indeed, when we compare the magnitude 
of the total exit rate to order $R^0$, shown
in Fig.~\ref{fig:radius_and_theoretical_rates} (b),
to the typical magnitude 
of the order-$R^0$ term Eq.~\eqref{eq:scenario1_Lagrangian} in Fig.~\ref{fig:radius_and_theoretical_rates} (c),
 we conclude that $\mathcal{L}_1^{\traj,(0)}$ only contributes less than 10\%
of the value of the total exit rate in subplot (b).

\textit{Scenario 2: Constant free-diffusion exit rate.}
From Eqs.~\eqref{eq:ratio_of_path_probabilities}, \eqref{eq:scenario1_free_diffusion}, 
we see that 
for constant tube radius the log-ratio of tube probabilities in general diverges 
as $R_0 \rightarrow 0$.
As discussed in Sect.~\ref{sec:asymptotic_tube_probabilities_N},
we can obtain a finite limit
 by choosing
 a path-dependent tube radius
$R(t) \equiv R(\traj(t)) \equiv R_0 \sqrt{D(\traj(t))/D_0}$\textcolor{black}{,
which for one-dimensional systems is equivalent
to defining the tube with respect to the metric induced by the diffusivity
tensor corresponding
 to the FP Eq.~\eqref{eq:ND_FP_eq_intro} \cite{ventsel_small_1970}.}
In this scenario, we obtain from Eqs.~\eqref{eq:axit_result_intro}, \eqref{eq:Lagrangian_1D}, that
\begin{align}
\label{eq:scenario2_free_diffusion}
\frac{
	\aexitfree^{\traj} (t)
	}{
	R(t)^2
	}
	&=
 \frac{\pi^2}{4} \frac{D_0}{R_0^2} \equiv const.
\\
\mathcal{L}_2^{\traj,(0)} &= 
 \frac{1}{4D} \left( \dot{\traj} - a + \frac{1}{2}\partial_x D\right)^2
+\frac{1}{2} \partial_x a - \frac{1}{4} \frac{\partial_x D}{D} a
\nonumber \\&  \quad 
\label{eq:scenario2_Lagrangian}
- \frac{\pi^2 - 2}{16} \frac{ \left( \partial_x D\right)^2}{D}
+ \frac{1}{4} \left( \frac{ \pi^2}{6} - 1 \right) \partial_x^2 D,
\end{align}
where the subscript 2 indicates that this is the second scenario we consider.
By construction, the free-diffusion exit rate Eq.~\eqref{eq:scenario2_free_diffusion}
is now independent of the path and  constant as a function of time.
In Fig.~\ref{fig:radius_and_theoretical_rates} (a),
we show the time-dependent tube radius $R(t)$ 
 for $R_0/L = 0.1$ and the 
  example system Eqs.~\eqref{eq:D_example}, 
  \eqref{eq:drift_potential},
\eqref{eq:reference_path}.
Figure \ref{fig:radius_and_theoretical_rates} (b)
 clearly shows that the total exit rate varies on a much smaller scale
as compared to the constant-radius exit rate from scenario 1,
which is because the dominant free-diffusion contribution Eq.~\eqref{eq:scenario2_free_diffusion}
 is now time-independent by design.
In Fig.~\ref{fig:radius_and_theoretical_rates} (c) we compare the
Lagrangians Eqs.~\eqref{eq:scenario1_Lagrangian}, \eqref{eq:scenario2_Lagrangian} 
for scenarios 1 and 2. 
While overall the Lagrangians are rather similar, they
deviate from each other 
for $t/\td \approx 0.5$,
when the path is close to the barrier top.
Because the two Lagrangians 
Eqs.~\eqref{eq:scenario1_Lagrangian}, \eqref{eq:scenario2_Lagrangian} 
have different 
functional forms, there is of course no a-priori reason to assume that they should lead
to identical curves.

\textit{Scenario 3: Stratonovich Lagrangian.}
Both Lagrangians Eqs.~\eqref{eq:scenario1_Lagrangian}, \eqref{eq:scenario2_Lagrangian},
are different from the Lagrangian for multiplicative noise originally derived 
by Stratonovich \cite{stratonovich_probability_1971}, which in our notation reads
\begin{align}
\mathcal{L}_{\mathrm{S}}^{\traj} &= 
 \frac{1}{4D} \left( \dot{\traj} - a + \frac{1}{2}\partial_x D\right)^2
+\frac{1}{2} \partial_x a - \frac{1}{4} \frac{\partial_x D}{D} a
\nonumber \\&  \quad 
\label{eq:scenario3_Lagrangian}
+ \frac{1}{8} \frac{ \left( \partial_x D\right)^2}{D}
- \frac{1}{4} \partial_x^2 D,
\end{align}
where the subscript S stands for Stratonovich.
We now demonstrate that the Lagrangian Eq.~\eqref{eq:scenario3_Lagrangian}
corresponds to an exit rate from a one-dimensional moving ball (i.e.~a time-dependent 
interval)
which is not centered at the path $\traj$, and which has a time-dependent radius that is
only to leading order identical to the radius
from scenario 2.
To derive Eq.~\eqref{eq:scenario3_Lagrangian}
from the 1D FP Eq.~\eqref{eq:1D_FP_eq}, 
 we introduce a new coordinate system
	$y \equiv \Phi(x)$
defined by
\begin{align}
\label{eq:coordinate_transformation_definition}
	\frac{\mathrm{d}\Phi}{\mathrm{d}x} &= \frac{1}{\sqrt{D(x)/D_0}}.
\end{align}
Transforming 
 Eq.~\eqref{eq:1D_FP_eq} to the $y$-coordinate
 leads to \cite{risken_fokker-planck_1984}
 $\partial_t{P}_{Y} = - \partial_y \left( a_Y P_Y\right) + D_0 \partial_y^2 P_Y$,
where
	$P_Y(y,t) \equiv \sqrt{D(x)/D_0}P(x,t)$,
	$a_Y(y) \equiv 
	 \sqrt{{D_0}/{D(x)}}[a(x) - (\partial_x D)(x)/2]$, with $x = \Phi^{-1}(y)$.
The coordinate transformation Eq.~\eqref{eq:coordinate_transformation_definition}  locally 
compresses
space
 where the diffusivity is small, and locally stretches 
 space where
the diffusivity is large, resulting
in a constant diffusivity $D_0$ with respect to
 the $y$-coordinate, as was remarked by Ito \cite{ito_probabilistic_1978}.
(As has been emphasized before \cite{de_pirey_path_2023}, 
a coordinate transformation that flattens the diffusivity profile is only guaranteed
to exist for one-dimensional multiplicative-noise systems.)
The Stratonovich Lagrangian now follows by considering
a  tube with constant radius $R_0$ 
in the $y$-coordinate
around the path
 $\traj_Y(t) \equiv \Phi(\traj(t))$.
Since the diffusivity is constant in this coordinate system,
the theory for stochastic dynamics with additive noise is applicable,
for which the first correction to freely-diffusive exit from the tube is
given by the OM Lagrangian Eq.~\eqref{eq:OnsagerMachlupLagrangian} \cite{kappler_stochastic_2020}
evaluated using diffusivity $D_0$, drift $a_Y$, and the path $\traj_Y$.
Expressing the
resulting exit rate back in the original $x$-coordinates yields
\begin{align}
\label{eq:scenario3_aexit}
	\alpha_{R_0}^{\traj} &= \frac{\pi^2}{4} \frac{D_0}{R_0^2} + \mathcal{L}_{\mathrm{S}}^{\traj} + \mathcal{O}(R_0^2)
\end{align}
with the Stratonovich Lagrangian Eq.~\eqref{eq:scenario3_Lagrangian}.
In summary, $\mathcal{L}_{\mathrm{S}}^{\traj}$ is obtained by performing a
nonlinear coordinate transformation $\Phi$
such that the diffusivity is constant in the new coordinates,
then considering a constant-radius tube in the new coordinates, and
finally expressing the resulting exit rate in terms of the original coordinates.
Importantly, a tube centered around $\traj_Y(t) \equiv \Phi(\traj(t))$ in the $y$-coordinate 
in general does  not
 correspond to a tube centered around $\traj(t)$ in the $x$-coordinate.
More explicitly, at time $t$ a tube of radius $R_0$ in the $y$-coordinate
is in the $x$-coordinate bounded by the two points
	$\traj_{\pm}(t) \equiv \Phi^{-1}\left( \Phi(\traj(t)) \pm R_0 \right)$.
The center 
$	\traj_{\mathrm{S}}(t) \equiv \left( \traj_{+}(t) + \traj_{-}(t)\right)/2 $
and radius 
$R_{\mathrm{S}}(t) \equiv \left( \traj_{+}(t) - \traj_{-}(t)\right)/2$
of this one-dimensional ball  are given by 
\begin{align}
\label{eq:strat_center}
	\traj_{\mathrm{S}}(t) 
	&= \traj(t) + \frac{1}{4}\frac{\partial_x D|_{x = \traj(t)}}{D_0} R_0^2 + \mathcal{O}(R_0^4),
	\\
	\label{eq:scenario3_R}
	R_{\mathrm{S}}(t) 
	&=\left. \left[ \sqrt{ \frac{D}{D_0}} R_0 + \frac{1}{12} \frac{\partial_x^2 D}{D_0} \sqrt{ \frac{D}{D_0}} R_0^3 
	\right]\right|_{x = \traj(t)}+ \mathcal{O}(R_0^5).
\end{align}
Equations \eqref{eq:strat_center},
\eqref{eq:scenario3_R},  are to leading order identical to 
tube center and radius from scenario 2, but
contain additional higher-order terms.

We return to the example system
Eqs.~\eqref{eq:D_example}, 
  \eqref{eq:drift_potential},
\eqref{eq:reference_path},
and consider a tube of constant radius $R_0/L = 0.1$ in the $y$-coordinate.
In Fig.~\ref{fig:radius_and_theoretical_rates} (a) we show
$R_{\pm}(t) \equiv \pm ( \traj_{\pm}(t) - \traj(t))$, which is
the
distance 
from either of the two tube boundaries
to the path $\traj$
in the $x$-coordinate.
The two curves $R_{\pm}$ clearly disagree with each other, showing that the
tube is
in the $x$-coordinate not centered at $\traj$.
Both $R_{\pm}$ behave similar to the tube radius from scenario 2,
which according to Eq.~\eqref{eq:scenario3_R} is 
 their leading order behavior.
Indeed, also in Fig.~\ref{fig:radius_and_theoretical_rates} (b),
where we show the total exit rate Eq.~\eqref{eq:scenario3_aexit}
to order $R^0$, the exit rates from scenarios 2 and 3 look very similar.
However, subtracting
from the exit rate the free-diffusion contributions, which 
according to Eqs.~\eqref{eq:scenario2_free_diffusion},
 \eqref{eq:scenario3_aexit}, are equal, 
 we observe in Fig.~\ref{fig:radius_and_theoretical_rates} (c)
that the order-$R^0$ Lagrangians from scenario 2 and 3, 
Eqs.~\eqref{eq:scenario2_Lagrangian}, \eqref{eq:scenario3_Lagrangian},
 clearly deviate from each other, most 
prominently at the beginning and the end of the path,
 i.e.~for $t/\td \lesssim 0.3$ and $t/\td \gtrsim 0.7$.

We now consider ratios of tube probabilities for two paths $\traj$, $\trajTwo$ for scenario 3.
By construction, the free-diffusion exit rates are equal for any two paths, so that
we obtain
\begin{align}
\label{eq:scenario3_ratio_limit}
\lim_{R_0\rightarrow 0}\ln \frac{\Ptube(\tfinal)}{\PtubeTwo(\tfinal)} &= - \int_{\tinitial}^{\tfinal}\mathrm{d}s \,
 \left[ \mathcal{L}_{\mathrm{S}}^{\traj}(s) - \mathcal{L}_{\mathrm{S}}^{\trajTwo}(s)\right].
\end{align}
We emphasize that while in both scenarios 2 and 3 the ratio of tube probabilities is well-defined
in the limit $R_0 \rightarrow 0$, the resulting stochastic Lagrangians differ.
This highlights that limiting ratios of tube probabilities, and also the Lagrangians associated with them,
 depend on the detailed nature of the  tube.
However, if we consider for $\trajTwo$ the reverse of $\traj$, i.e.~$\trajTwo = \overset{\leftarrow}{\traj}$, 
then from Eq.~\eqref{eq:scenario3_ratio_limit} we recover
Eq.~\eqref{eq:1D_limit_radius_zero_reverse} without the boundary term,
i.e.~we obtain the usual multiplicative-noise pathwise entropy production.

\subsection{Most probable tube}
\label{sec:most_probable_tube}

\begin{figure}[ht]
\centering \includegraphics[width=\columnwidth]{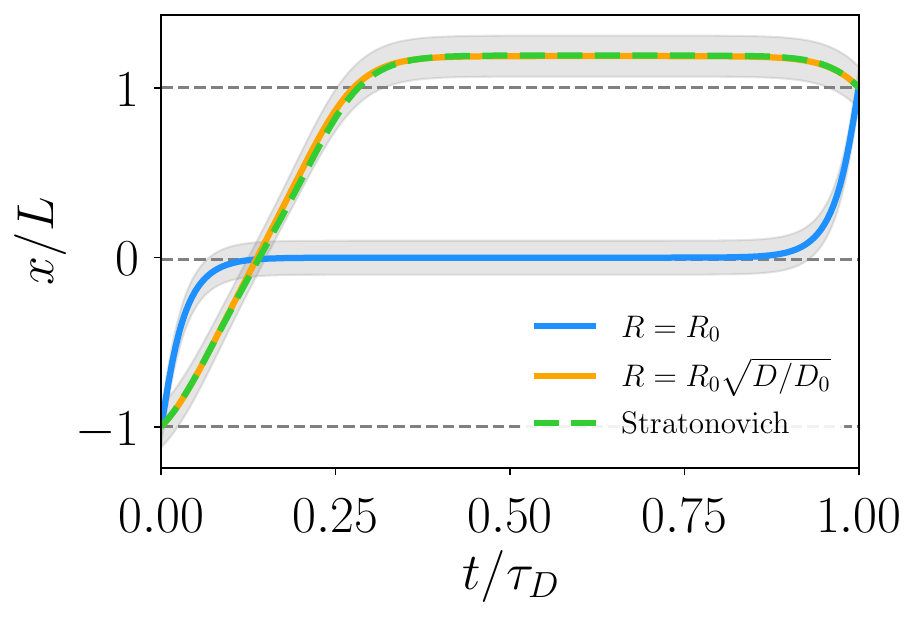} 
\caption{ \label{fig:instantons_0} 
Most probable tubes for the three scenarios considered 
 in Sect.~\ref{sec:results_for_one_dimensional_system},
 with force and diffusivity given by Eqs.~\eqref{eq:D_example}, \eqref{eq:drift_potential}.
We minimize
Eq.~\eqref{eq:def_most_probable_tube}
 using the same algorithm as used for functional minimization in Ref.~\cite{gladrow_experimental_2021}.
More explicitly, we approximate the path in Eq.~\eqref{eq:def_most_probable_tube}
 by a finite number of $40$ modes,
and solve the resulting finite-dimensional minimization problem for the mode coefficients
 using a standard algorithm \cite{hansen_cma-espycma_2019}; 
 see Ref.~\cite{gladrow_experimental_2021} for more details.
For the blue and orange solid lines, the 
minimization is carried out for the temporal integral 
over the exit rate Eq.~\eqref{eq:axit_result_intro} to order $R^0$,
once
with a constant tube radius $R_0/L = 0.1$ (blue solid line) and once with 
a path-dependent tube radius $R(t) \equiv R_0 \sqrt{ D(\traj(t))/D_0}$ where $R_0/L = 0.1$ (orange solid line);
both tubes are indicated as gray shaded area.
For the blue dashed line, the temporal integral over the exit rate Eq.~\eqref{eq:scenario3_aexit}
is minimized using $R_0/L = 0.1$.
Initial and final position of the paths are shown as horizontal dashed lines.
}
\end{figure}

We now consider the most probable tube
for the
  system Eqs.~\eqref{eq:D_example}, \eqref{eq:drift_potential} 
and the three scenarios discussed in the previous section.
For each scenario, we evaluate the most probable tube for a barrier-crossing transition 
from $\traj(0) = \xl \approx -L$ to $\traj(\tfinal) = \xr \approx L$ in one unit of the diffusive time scale, $\tfinal = \td$.
We minimize the action functional Eq.~\eqref{eq:def_most_probable_tube} to order $R^0$ (inclusive)
for $R_0/L = 0.1$, and
in Fig.~\ref{fig:instantons_0} compare the resulting MPT centers $\traj^*$.

For constant tube radius $R(t) \equiv R_0/L = 0.1$, the most probable reference
path $\traj^*$ (blue solid line)
remains on the barrier top $x = 0$ for most of the transition time.
This is because the action is, for the small radius considered here, dominated by the 
free-diffusion exit rate Eq.~\eqref{eq:scenario1_free_diffusion}, 
which is proportional to the diffusivity.
Because the diffusivity profile Eq.~\eqref{eq:D_example} 
features a local minimum at $x/L = 0$, the free-diffusion exit rate is minimal there.
Notably, it follows that for tubes of small constant radius 
the most probable tube is only weakly influenced by the Lagrangian
Eq.~\eqref{eq:scenario1_Lagrangian}, and in particular is
dominated by the diffusivity profile $D(x)$ as 
 compared to the drift $a(x)$.
This very effect is also observed in the experimental data analyzed
 in Ref.~\cite{accompanying_paper_short}.

For scenario 2, where $R(t) \equiv R_0 \sqrt{ D(\traj(t))/D_0}$, 
 the free-diffusion exit rate is  constant,
 so that the extremum of the action is determined by the subleading-order contribution
 Eq.~\eqref{eq:scenario2_Lagrangian}.
The resulting MPT shown in Fig.~\ref{fig:instantons_0} 
crosses the barrier rather quickly without stopping at the barrier top,
and then
 remains within the target potential well most of the
time;
this is in sharp contrast to the
 constant-radius result.
Indeed, the most probable tube observed here
is qualitatively similar to the most probable barrier-crossing paths
for double well systems with constant diffusivity 
\cite{adib_stochastic_2008,gladrow_experimental_2021}, i.e.~to the 
MPPs which follows from 
the OM Lagrangian Eq.~\eqref{eq:OnsagerMachlupLagrangian}.

For the Stratonovich scenario the most probable tube 
 is obtained by minimizing the integrated exit rate Eq.~\eqref{eq:scenario3_aexit} as a functional of $\traj$.
We show the resulting
 most probable reference path $\traj^*$
in Fig.~\ref{fig:instantons_0} as green dashed line.
We observe that this MPT center is almost identical to the result
 for constant free-diffusion exit rate.
This is consistent with the facts  
that in the Stratonovich scenario the
free-diffusion exit rate is also independent of the path, c.f.~Eq.~\eqref{eq:scenario3_aexit},
and that 
the $R^0$-order terms for both scenarios 2 and 3
yielded similar results also in the previous subsection, see Fig.~\ref{fig:radius_and_theoretical_rates} (c).

\begin{figure}[ht]
\centering \includegraphics[width=\columnwidth]{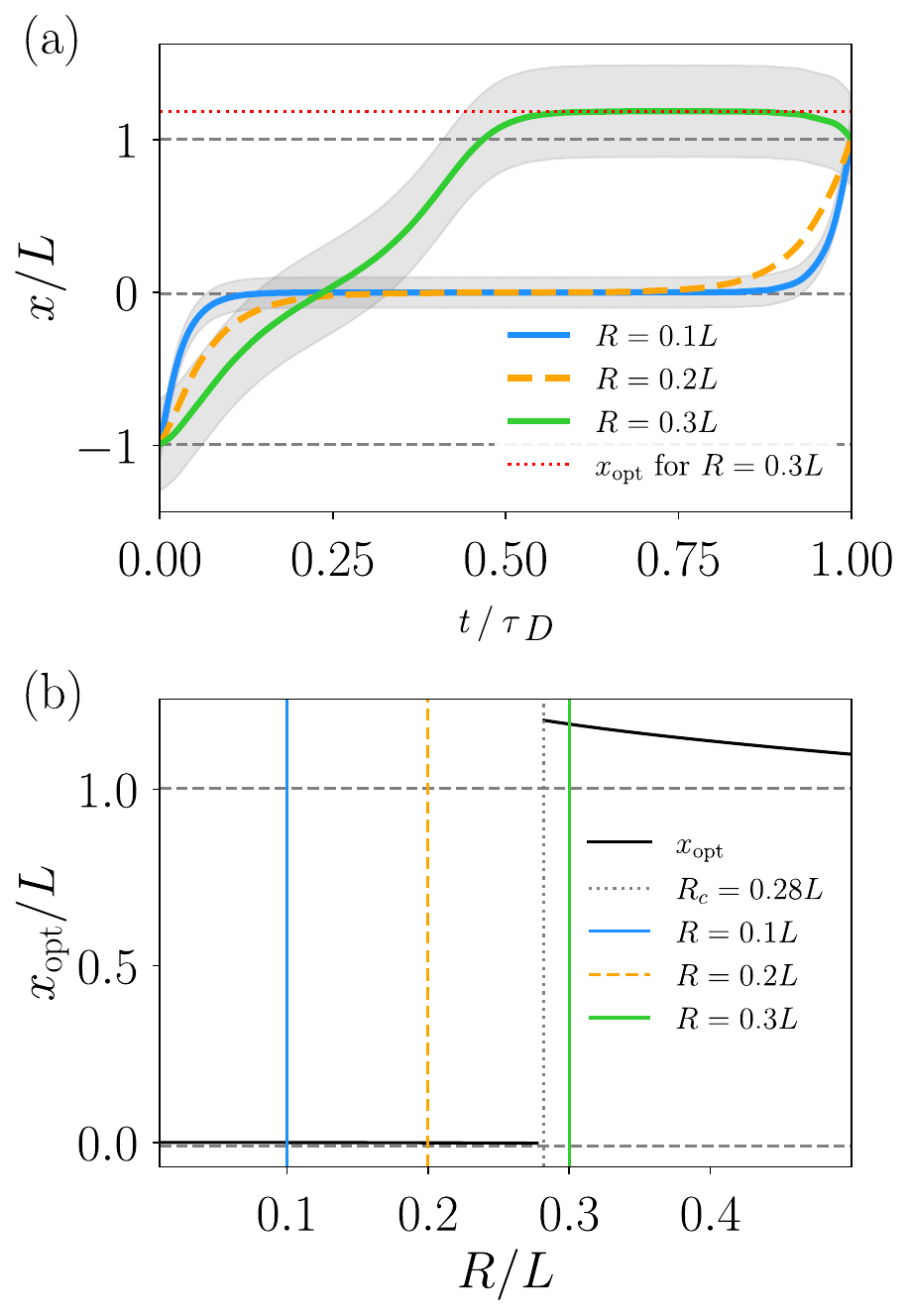} 
\caption{ \label{fig:radius_dependence_of_instanton} 
(a) 
Most probable constant-radius tube reference path for various tube radii.
We minimize
Eq.~\eqref{eq:def_most_probable_tube}
 using the algorithm 
  described in App.~\ref{app:functional_minimization_algorithm}.
The minimization is carried out using the temporal integral
over the exit rate Eq.~\eqref{eq:axit_result_intro} to order $R^2$,
with force and diffusivity given by Eqs.~\eqref{eq:D_example}, \eqref{eq:drift_potential}.
We show the resulting most probable reference path $\traj^*$  for
 a constant tube radius $R_0/L = 0.1$ (blue solid line), $R_0/L = 0.2$ (orange dashed line),
and $R_0/L = 0.3$ (green solid line).
For the smallest and largest tube radius, we indicate the tube as gray shaded area.
Initial and final position of the paths are shown as horizontal dashed lines.
The red horizontal dotted line denotes the value of the function $x_{\mathrm{opt}}$ 
at $R = 0.3\,L$.
(b)
The black solid line shows the location $x_{\mathrm{opt}}$ such that the exit rate from
 the interval of width $2R$ is minimal, as defined in
Eq.~\eqref{eq:x_opt_resting}.
The vertical dotted line shows the crossover radius $R_{c} \approx 0.28\,L$
at which $x_{\mathrm{opt}}$ is discontinuous as a function of the radius $R$.
The colored vertical lines denote the three values of the tube radius considered
in subplot (a).
The gray horizontal dashed lines denote the local extrema of the
potential shown in Fig.~\ref{fig:potential_and_trajectory} (b).
}
\end{figure}

To close this section, we consider the radius dependence of the
most probable tube Eq.~\eqref{eq:def_most_probable_tube};
this discussion parallels the corresponding discussion
in the accompanying Ref.~\cite{accompanying_paper_short}.
We consider a tube of constant radius $R \equiv R_0 = const.$ for 
the example system
Eqs.~\eqref{eq:D_example}, \eqref{eq:drift_potential}.
As in
Fig.~\ref{fig:instantons_0}, 
we consider paths that move from $x=\xl \approx-L$ to $x=\xr \approx L$ 
during a time $\tfinal = \td$.
We minimize  
Eq.~\eqref{eq:def_most_probable_tube}, evaluated using Eq.~\eqref{eq:axit_result_intro} 
to order $R^2$ (inclusive),
for each of the constant-radius tubes $R_0/L = 0.1$, $0.2$, $0.3$.
We show the resulting most probable reference paths
in Fig.~\ref{fig:radius_dependence_of_instanton} (a).  
For $R_0/L = 0.1$ we obtain the same path $\traj^*$ as shown in Fig.~\ref{fig:instantons_0},
which shows that the quadratic term in the exit rate is irrelevant for
this small tube radius; see App.~\ref{app:one_dim} for more details.
As discussed in the context of Fig.~\ref{fig:instantons_0},
for $R_0/L = 0.1$ the MPT center remains on the barrier top for
 most the transition time because the exit rate
is  for small radius dominated by the free-diffusion contribution Eq.~\eqref{eq:scenario1_free_diffusion}.
While for $R_0/L = 0.2$ the most probable reference path $\traj^*$  
also rests at the barrier top for most of the transition time, 
it stays there for a shorter duration as compared to the $R_0/L = 0.1$ result;
this indicates that for $R_0/L = 0.2$ the free-diffusion exit rate is already less dominant.
For $R_0/L = 0.3$, the path $\traj^*$ is completely different from
its smaller-radius counterparts.
Now, the most probable reference path immediately crosses over the potential barrier
without stopping, and rests close to the potential 
minimum $\xr$
for
most of the transition time.
The behavior of $\traj^*$ is now more reminiscent of the 
most probable tube for the constant free-diffusion 
exit rate scenario from
Fig.~\ref{fig:instantons_0}.

The reason for the crossover that we observe in
Fig.~\ref{fig:radius_dependence_of_instanton} (a)
is the radius-dependent competition between the terms in the perturbation series
Eq.~\eqref{eq:axit_result_intro}.
To investigate this competition further we now consider a constant path at a point $x$
with constant radius $R \equiv R_0 \equiv const.$,
and ask for which choice of $x$ the exit rate is minimal, i.e.~we consider
the function
\begin{equation}
\label{eq:x_opt_resting}
x_{\mathrm{opt}}(R) = \underset{x}{\mathrm{argmin}} \,\alpha_R^{\traj_x},
\end{equation}
where $\traj_x(t) \equiv x$ is the path resting at $x$.
For any value of $R$ the point $x_{\mathrm{opt}}(R)$ is
is the most stable point in the system, in the sense that
the steady-state exit rate 
from
the interval $[ x_{\mathrm{opt}}- R, x_{\mathrm{opt}} + R]$ 
is the minimal exit rate achievable for any
 interval of width $2R$ in the system.
In Fig.~\ref{fig:radius_dependence_of_instanton} (b)
we show Eq.~\eqref{eq:x_opt_resting} 
as a function of $R$.
We observe that for small radius, the optimal resting path 
is located at $x \approx 0$.
This is because for small radius
the exit rate Eq.~\eqref{eq:axit_result_intro}
is dominated by the free-diffusion term Eq.~\eqref{eq:scenario1_free_diffusion}, 
which is minimal at $x = 0$.
As the radius is increased, the terms at order $R^0$, $R^2$ in Eq.~\eqref{eq:axit_result_intro}
become more important.
At  the crossover radius $R_c \approx 0.28\,L$ the optimal resting point $x_{\mathrm{opt}}$ is
 discontinuous
as a function of $R$,
and jumps from $x\approx 0$ to
the vicinity of the local minimum $\xr$.
Thus, for $R > R_c$ the confinement effects of the potential landscape around $\xr$
 outweigh
the benefit of the small diffusivity at $x =0$.
This explains the behavior of the MPT centers from
Fig.~\ref{fig:radius_dependence_of_instanton} (a),
which according to Fig.~\ref{fig:radius_dependence_of_instanton} (b)
for every radius considered rest at the 
respective
most stable ball in the system.
We note that the optimal resting position is never close to the minimum $\xl$;
the minimum $\xr$ is preferred because of
 the linear term in the potential Eq.~\eqref{eq:drift_potential}.
In fact, 
we have included the
symmetry-breaking term precisely
so that one of the two minima is preferred over the other.

This example shows that the most probable pathway for a transition can depend
significantly on the tube radius, i.e.~on how much deviation 
from the reference path
one tolerates.

\section{Conclusions}
\label{sec:summary_and_conclusions}

In this work we present a general theory for the sojourn probability, which is
the probability for a diffusive trajectory
to remain within a tube of small but finite time-dependent radius $R(t)$ around 
 a continuous reference path $\trajvec(t)$.
We focus on the case of
 $N$-dimensional Langevin dynamics with multiplicative noise and full-rank diffusivity matrix, 
a tube defined using the standard Euclidean norm, 
and twice continuously differentiable reference paths.
For this scenario we derive an expansion in powers of the tube radius
 for the instantaneous exit rate at which 
 stochastic trajectories first leave a small-but-finite radius tube.
Based on this exit rate, we discuss the vanishing-radius limit
for ratios of sojourn probabilities for pairs of reference paths.
We show that while in general such limiting ratios are either zero or divergent, 
for a pair of forward and reverse path they are finite.
For the special case of a one-dimensional system, $N=1$, 
we derive explicit expressions for the exit rate in terms of the drift and
diffusivity,  
consider several choices for the time-dependent tube radius,
and illustrate our results with an explicit numerical example.
The Lagrangian Eq.~\eqref{eq:Lagrangian_1D} we 
derive for one-dimensional Langevin dynamics
is different from Lagrangians found in the literature \cite{stratonovich_probability_1971,
durr_onsager-machlup_1978,
cugliandolo_building_2019,stratonovich_probability_1971,horsthemke_onsager-machlup_1975,
fujita_onsager-machlup_1982,ikeda_stochastic_1989,
ito_probabilistic_1978,williams_probability_1981},
and has the advantage of being directly related to an observable exit rate.
For our one-dimensional example system, we furthermore illustrate how
the most probable tube depends on both the choice of time-dependence of 
the tube radius, as well as the size of the tube.
Our results have several important consequences, from both a mathematical and physical
point of view.

The exit rate we derive is for small radius dominated by a free-diffusion contribution,
which for additive noise
  is independent of the reference path.
In this case  limiting ratios of sojourn probabilities for constant-radius tubes 
probe subleading-order terms of the exit rate, and 
can be used to define the 
stochastic action \cite{stratonovich_probability_1971,
fujita_onsager-machlup_1982,ikeda_stochastic_1989,
ito_probabilistic_1978,williams_probability_1981,
durr_onsager-machlup_1978,
horsthemke_onsager-machlup_1975,
kappler_stochastic_2020}.
For state-dependent noise, the local free-diffusion exit rate 
is also state-dependent, and the ratio of sojourn probabilities for two constant-radius tubes
is in general either zero or divergent in the limit of vanishing tube radius;
this means that one path is typically infinitely more likely than the other.
Our theory thus provides an intuitive physical picture as to why classical definitions of
 stochastic actions for additive-noise systems cannot be simply generalized to 
 systems with multiplicative noise \cite{durr_onsager-machlup_1978}.

Our work elucidates the geometry behind mathematical attempts
to obtain a finite limiting-ratio for pairs of sojourn probabilities \cite{stratonovich_probability_1971,williams_probability_1981,fujita_onsager-machlup_1982,ikeda_stochastic_1989}.
These works do not consider the sojourn probability to remain within a moving ball 
(defined with respect to 
the standard Euclidean norm)
centered at a reference path, but instead
the sojourn probability to remain within more complicated geometrical shapes, which are
not necessarily centered at the reference path.
\textcolor{black}{Since these definitions of the tubular neighborhood
use the diffusion tensor of the underlying stochastic dynamics,
this diffusion tensor needs to be inferred before   such finite-radius sojourn probabilities can be measured.
More so, from an experimental point of view it is more natural to simply consider a constant-radius
tube with respect to the metric of the ambient Euclidean space.}
The corresponding sojourn probability is for one-dimensional systems quantified by our
explicit results
 Eqs.~\eqref{eq:scenario1_free_diffusion}, \eqref{eq:scenario1_Lagrangian},
 which we compare to experimental measurements
 in the accompanying Ref.~\cite{accompanying_paper_short}.

While limiting ratios of sojourn probabilities for arbitrary pairs of paths do not lead to
finite results, a pair consisting of forward and reverse path does.
This implies that limits of sojourn probabilities can be used to quantify irreversibility 
along individual paths, and indeed for $N=1$ our results recover established
formulas for the path-wise
 entropy production \cite{seifert_entropy_2005,bo_functionals_2019,kappler_measurement_2022}.

Besides discussing the technical difficulties and ambiguities arising from 
trying to define vanishing-radius limits of sojourn probabilities for systems with multiplicative noise, 
our work focuses on considering finite-radius tubes. 
Our philosophy here is similar to that of a simplified version of our theory on additive
noise \cite{kappler_stochastic_2020}.
Namely, from a mathematical perspective, instead of trying to introduce a 
probability \textit{density} on the
infinite-dimensional space of all continuous paths, 
we evaluate the probability \textit{measure} induced on that space
by Langevin dynamics.
Because of this we do not need to consider limiting procedures in our theory, and indeed
it has been shown that for Langevin dynamics with multiplicative noise 
there is no canonical way of defining
a probability density on the space
of all continuous paths \cite{durr_onsager-machlup_1978}.

From a physical point of view, considering the finite-radius tubular ensemble is also reasonable:
The probability to observe a given individual trajectory vanishes, so that
it is not straightforward to quantify it in an experiment. 
The probability to observe any stochastic trajectory of the finite-radius tubular ensemble is positive, 
and hence is directly accessible in experiment,
simply by counting how many stochastic trajectories that started within the tube remain
until a later time \cite{kappler_stochastic_2020}.
Finite-radius tubes can thus be used to probe path-properties in experiment and simulation, 
and indeed for additive isotropic 
noise they have been used to infer both ratios of path probabilities \cite{gladrow_experimental_2021}
and the entropy production along individual paths \cite{kappler_measurement_2022}.
For multiplicative noise we infer finite-radius tube probabilities in the
accompanying Ref.~\cite{accompanying_paper_short}.

Our results demonstrate that the most probable tube depends sensitively on both the protocol
for the \textcolor{black}{state- or time-dependence} of the tube radius, and on the typical size of the tube.
Thus, because in practice there is typically a finite amount of deviation from a reference path
 one is willing to tolerate, 
considering the single most probable path in general does not yield physically relevant results.
The concept of the most probable tube will be useful for understanding
in more depth the properties of transition paths 
\cite{dykman_optimal_1992,luchinsky_analogue_1998,chan_paths_2008,
lehmann_activated_2003,
schorlepp_gelfandyaglom_2021,
kikuchi_diffusivity_2022,
schorlepp_symmetries_2023},
for example by investigating how a small \textcolor{black}{state- or time-dependent}
tube radius can be chosen so as to capture as many finite-temperature transition paths as possible.

Another interesting direction for future research is to consider the ratio of sojourn probabilities
for forward/reverse path pairs also at finite tube radius, and to relate the resulting expression to
the path integral of the single-trajectory entropy production over all stochastic trajectories in the 
corresponding tubular ensemble.
This will yield an experimentally relevant
 generalization of the pathwise entropy production \cite{maes_time-reversal_2003,
seifert_entropy_2005,seifert_stochastic_2012,kappler_measurement_2022} to tubes.

To date, for multiplicative noise and dimension $N \geq  2$,
no explicit representation in terms of $\avec$, $\Dmat$ is available for 
the exit rate Eq.~\eqref{eq:axit_result_intro} from a tube with small-but-finite constant radius,
defined via the standard Euclidean metric.
Since this exit rate is arguably the most straightforward experimental observable 
for quantifying
the probability of a given pathway,
an important next step will be calculating
 explicit expressions for the exit rate, in terms of $\avec$, $\Dmat$,
  also for dimensions $N \geq 2$.
From our results for one-dimensional systems, it is expected
that the order-$R^0$ contribution to the resulting exit rate will 
be different from 
  the Stratonovich Lagrangian \cite{stratonovich_probability_1971}.
  
Similarly, for dimensions $N \geq 2$, it will be interesting to derive a theory
for stochastic dynamics with a diffusivity tensor that is not full rank; 
this case is not covered by our approach here.
An important example system with a diffusivity tensor that is not full rank
 is given by  underdamped Langevin dynamics; there,
the degrees of freedom are the position $x$ and the velocity $v$ of a massive particle.
In this system the noise only acts on the velocity, so that the short-time dynamics of the tuple $(x,v)$ are not
as straightforward as in our present theory.

Beyond underdamped Langevin dynamics, the sojourn probability for a tube around a reaction 
coordinate is 
an experimentally relevant quantification of observable path probabilities for 
any kind of stochastic dynamics. It will therefore be interesting to compare tubular exit rates
to path integral actions also for e.g.~non-Markovian or active stochastic processes \cite{hanggi_path_1989,
hanggi_path_1993,
dabelow_irreversibility_2019}.
  
It will furthermore be interesting to relate our finite-noise theory 
to the low-noise theory of Freidlin and Wentzell \cite{ventsel_small_1970}.
They consider tubes for asymptotically small noise strength,
where the probability distribution on the space of all paths is concentrated 
around a single most probable tube center (called instanton in this context). 
At finite noise strength, on the other hand, the probability distribution on the space of paths has
a larger support, so that even the probability for the most probable tube at small but finite radius is typically
very small (we show this explicitly in Ref.~\cite{accompanying_paper_short}).
One should be able to observe a crossover from our theory
 to that of Freidlin and Wentzell by studying the
dependence of the MPT
on the typical diffusivity amplitude. 
More explicitly, upon decreasing the state-dependent diffusivity
while keeping the deterministic drift and the small radius constant, the MPT should cross over from
being dominated by the free-diffusion behavior we discuss in the present work, 
to being described by Freidlin-Wentzell theory.

In summary, our present work on sojourn probabilities for diffusive stochastic dynamics
provides a comprehensive and physical picture of the rather technical literature on
path probabilities for systems with state-dependent noise, relates the
concept of path probabilities for such systems to measurement, 
and in particular for the first time quantifies the probability for a stochastic trajectory
to remain within a constant-radius tube around a twice continuously differentiable reference path.

\begin{acknowledgments}

Work funded in part by the European Research Council under the Horizon 2020 Programme, ERC grant agreement number 740269, and by the Royal Society through grant RP1700.
J.~K.~acknowledges funding from the European Union's Horizon 2020 research and innovation programme under the Marie Sk{\l}odowska-Curie grant agreement No 101068745.

\end{acknowledgments}

\appendix

\section{Perturbative tubular exit rate for $N$-dimensional It\^o processes}
\label{app:N_dim_theory}

We here derive the expression Eq.~\eqref{eq:axit_result_intro} for
 the exit rate from 
a tube with time-dependent radius $R(t)$ around a path $\trajvec(t)$
for the It\^o-Langevin Eq.~\eqref{eq:Ito_process_definition}.
The present derivation generalizes the calculation in Ref.~\cite{kappler_stochastic_2020}
which considers additive isotropic noise
and a time-independent tube radius;
we here only highlight the differences to this previous derivation, and refer the reader 
to the reference for more details.
Throughout this appendix, we assume that $\trajvec$ is twice countinuously differentiable,
and use the standard Euclidean norm $||\xvec||_2 \equiv \sqrt{ x_1^2 + x_2^2 + \cdots + x_N^2}$
to define the tubular neighborhood of $\trajvec$.
To consider power series expansions in the time-dependent radius,
 we assume that $R(t)$ is of the form $R(t) = R_0 r(t)$ with a differentiable
 dimensionless function $r(t)$;
 any series expansion in powers of $R(t)$ actually refers to an expansion in powers of 
$R_0$.

\subsection{FP equation in dimensionless streaming coordinates.}
To solve Eq.~\eqref{eq:ND_FP_eq_intro} for a given reference path $\trajvec$, 
we transform to dimensionless coordinates $(\xvecDL, \tDL)$ 
with respect to which the domain of the FPE is time-independent.
We define the dimensionless coordinates as
\begin{align}
\tDL(t) & \equiv\frac{t}{\td}, 
\label{eq:def_dimensionless_coordinates}
& \xvecDL(\xvec,t) & \equiv\frac{\xvec-\trajvec(t)}{R(t)},
\end{align}
where we define a diffusive time scale $\td = L^2/D_0$, 
with $D_0$ a typical diffusive scale of the system and $L$ a typical
length scale.
The definition Eq.~\eqref{eq:def_dimensionless_coordinates} generalizes the 
coordinates used in Ref.~\cite{kappler_stochastic_2020}, 
where only a time-independent tube radius was considered.
With respect to the coordinates $(\xvecDL,\tDL)$, the domain for the FPE is the unit ball,
$\xvecDL \in \BDL \equiv \{~||\xvecDL||_2 \leq 1~\}$.
At time $\tDL$, the absorbing boundary conditions 
are then given by
$\PtubeDL(\xDLvec,\tDL)=0$
for $||\xDLvec||_2=1$,
where the dimensionless density is defined as 
$\PtubeDL(\xvecDL,\tDL)  \equiv R(t)^{N}\Ptube\left(\xvec,t\right)$.
Casting the FP Eq.~\eqref{eq:ND_FP_eq_intro} inside the tube 
in dimensionless form yields
\begin{align}
\epsilonDL^{2}\,\partial_{\tDL}\PtubeDL & =\LFPDLapp\PtubeDL,\label{eq:dimensionless_Ndim_FPE}
\end{align}
where we define the dimensionless apparent FP operator
\begin{align}
\LFPDLapp\PtubeDL &\equiv
-\epsilonDL\, \partialDL_i \left(\aDL_{\mathrm{app},i} \PtubeDL \right)
\label{eq:dimensionless_Ndim_FPO}
+
\partialDL_i \partialDL_j \left( \DDL_{ij} \PtubeDL \right)
\end{align}
with 
$\avecDL(\xvecDL,\tDL)  \equiv \td \avec\left(\vecx,t\right)/L$,
$\DmatDL(\xvecDL,\tDL)  \equiv \Dmat\left(\vecx,t\right)/D_0$,
$\vectrajDL(\tDL)  \equiv \vectraj(t)/L$,
$\tilde{\nabla}_{j}\equiv \partial/\partial \xDL_j = R \,\partial/\partial x_j$,
$\epsilonDL(\tDL)\equiv R(t)/L$,
$\avecDL_{\mathrm{app}}=
\avecDL 
-\vectrajdotDL
- \dot{\epsilonDL}\, \xvecDL$.
Throughout this work, a dot 
on a dimensionless quantity, as indicated by a tilde,
denotes a derivative with respect to dimensionless time $\tDL$,
and a dot on a quantity in physical units indicates
 a derivative with respect to $t$; 
 with this convention we have
 for example that $\vectrajdotDL = \vectrajdot\, \td/L$.
For a system with constant isotropic diffusivity, $\Dmat \equiv D_0 \underline{\mathbb{1}}$, and
a time-independent tube radius $R(t) \equiv R_0 = const.$, the dimensionless apparent
FP operator Eq.~\eqref{eq:dimensionless_Ndim_FPO} reduces to its counterpart 
in Ref.~\cite{kappler_stochastic_2020}.

\subsection{Perturbative FP propagator.}
An approximate propagator 
(i.e.~transition probability density from an initial to a final point)
for the dimensionless FP Eq.~\eqref{eq:dimensionless_Ndim_FPE} can be
derived for small tube radius
 by first projecting the FPE onto its instantaneous
eigenbasis, and subsequently solving the resulting projected equation
using an approach similar to time-dependent
perturbation theory in quantum mechanics \cite{ballentine_quantum_2010}.
The derivation, which is given in detail in Ref.~\cite{kappler_stochastic_2020}, 
 is also applicable in the present scenario.
There are, however, two important differences between the previous derivation and the current scenario.
First, in the present work we consider a time-dependent dimensionless tube radius $\epsilonDL$,
whereas in Ref.~\cite{kappler_stochastic_2020} the tube radius was assumed constant.
Second, because the  diffusivity matrix components 
$\DDL_{ij}(\xDL)$ in Eq.~\eqref{eq:dimensionless_Ndim_FPO}
are state-dependent, 
the spectrum of the dimensionless FPE in general depends on
$\tDL$ also to lowest order in $\epsilonDL$ (with the exception of the 
 one-dimensional case $N=1$, for which 
 we show in App.~\ref{app:one_dim} that the spectrum is to lowest order still independent 
of $\tDL$).
By contrast,
 in Ref.~\cite{kappler_stochastic_2020}, the spectrum
was to lowest order independent of $\tDL$. 
Thus, while 
in this previous work mode-coupling effects were only relevant at order $\epsilonDL^3$,
in the present work they can become relevant already at order $\epsilonDL^2$
for dimensions $N \geq 2$.
Taking these two differences 
  into account, the
derivation of a perturbative propagator
 for Eq.~\eqref{eq:dimensionless_Ndim_FPE} is
  carried out as in Ref.~\cite{kappler_stochastic_2020},
resulting in
\begin{align}
\label{eq:perturbative_Ndim_propagator}
	\PtubeDL
	&(\xvecDL,\tDL~\setbar~\xveczeroDL,\tzeroDL) =
	 \exp\left[ - \int_{\tinitialDL}^{\tDL}\mathrm{d}\tDLdummy~\frac{\Lone(\tDLdummy)}{\epsilonDL^2(\tDLdummy)} \right] 
	\\ \nonumber
	 &
	\times
	 \frac{1}{\reqDL(\xzeroDLvec,\tinitialDL) {\langle \efone,\efone \rangle|_{\tinitialDL}} } 
	 \\
	 & \times
 \left[ \efone(\xvecDL,\tDL) - \sum_{m=2}^{\infty}
    \nonumber
   \left.\frac{\epsilonDL^2(\tDL)}{\Lmone(\tDL)} 
   \frac{\langle \efm,\efonedot\rangle}{\langle \efm,\efm\rangle}\right|_{\tDL}
   \efm(\xvecDL,\tDL) \right]
   \\ & \times
 \left[ \efone(\xveczeroDL,\tinitialDL) -\sum_{n=2}^{\infty}
 \nonumber
   \left.
   \frac{\epsilonDL^2(\tinitialDL) }{\Lnone(\tinitialDL)}
   \frac{\langle \efone,\efndot\rangle}{\langle \efn,\efn\rangle}\right|_{\tinitialDL}
   \efn(\xveczeroDL,\tinitialDL) \right] 
   \\
	 \nonumber
   	    &+ \mathcal{O}(\epsilonDL^k),
\end{align}
where $k = 4$ for $N \geq 2$ and $k = 6$ for $N=1$,
where the instantaneous 
eigenvalues $-\evn$ and eigenfunctions $\efn$ of the dimensionless apparent FP operator are defined 
via
\begin{align}
\LFPDLapp(\tDL)\efn(\xvecDL,\tDL) &
\label{eq:dimensionless_FPO_eigenvalue_equation}
=-\evn(\tDL)\efn(\xvecDL,\tDL)
\end{align}
with absorbing boundary conditions $\efn(\xvecDL,\tDL) = 0$ for $||\xvecDL||_2 = 1$,
and with
$\Ln  \equiv\evn+\epsilonDL^{2}\langle\efn,\efndot\rangle/\langle\efn,\efn\rangle$
and
$\Lmn  \equiv\Lm-\Ln$.
The time-dependent inner product of two functions $g$, $h$ is defined as
\begin{align}
\langle g,h\rangle&\equiv
\label{eq:inner_product_definition}
\int_{\BDL}\mathrm{d}^{N}\xvecDL~g(\xvecDL)h(\xvecDL)/\reqDL(\xvecDL,\tDL),
\end{align}
where  $\reqDL$ is the instantaneous reflecting-boundary
 steady state corresponding to Eq.~\eqref{eq:dimensionless_Ndim_FPO} inside the unit ball $\BDL$ \cite{kappler_stochastic_2020}.
The solution Eq.~\eqref{eq:perturbative_Ndim_propagator} is valid after the transient decay of the initial condition,
i.e.~for 
\begin{equation}
\label{eq:relaxation_time}
\tDL- \tzeroDL \gtrsim \trelDL \equiv \frac{\epsilonDL^{2}(\tzeroDL)}{ \Delta\Lambda_{21}(\tzeroDL)}.
\end{equation}
Note that for the 
special case of a one-dimensional system,
we in App.~\ref{app:transient_decay}
 give an expression for the perturbative propagator Eq.~\eqref{eq:perturbative_Ndim_propagator}
 that is also valid for shorter times.

The approximate propagator Eq.~\eqref{eq:perturbative_Ndim_propagator} neglects both terms that are exponentially smaller
than the expression Eq.~\eqref{eq:perturbative_Ndim_propagator}, 
as well as terms that are at least of order $\epsilonDL^4$.
In contrast to its counterpart in Ref.~\cite{kappler_stochastic_2020}, the propagator 
Eq.~\eqref{eq:perturbative_Ndim_propagator} i) features a time-dependent $\epsilonDL$,
and ii) is valid only to order $\epsilonDL^3$ for dimensions $N \geq 2$.

\subsection{Perturbative exit rate.}
In dimensionless form, the exit rate Eq.~\eqref{eq:inst_exit_rate} is given by
\begin{align}
\aexitDL_{R}^{\trajvec}(\tDL)
\label{eq:exit_rate_DL}
&\equiv\td\,\aexit(t)=-\frac{\PtubedotDL(\tDL)}{\PtubeDL(\tDL)},
\end{align}
where 
$\PtubeDL(\tDL)= \int_{\BDL}\mathrm{d}^N\xvecDL\,\PtubeDL(\xvecDL,\tDL)$
is the dimensionless survival probability up to time $\tDL$.
Similar to Ref.~\cite{kappler_stochastic_2020}, 
we evaluate Eq.~\eqref{eq:exit_rate_DL} using the
 approximate propagator Eq.~\eqref{eq:perturbative_Ndim_propagator}.
The resulting expression, which is valid for $\tDL- \tzeroDL \gtrsim \trelDL$, is independent of
the initial condition and given as a power series in $\epsilonDL$ by
\begin{align}
\aexitDL_{R}^{\trajvec} &=
\label{eq:dimensionless_N_dimesional_exit_rate_power_series}
\aexitfreeDL^{\trajvec}
+\aexitDL^{\trajvec,(0)}
+\epsilonDL^{2}\aexitDL^{\trajvec,(2)}
+\mathcal{O}(\epsilonDL^{4}),
\end{align}
where the individual terms are expressed in terms of the instantaneous spectrum as
\begin{align}
\aexitfreeDL^{\trajvec} & =\frac{\evone^{(0)}}{\epsilonDL^{2}},\label{eq:dimensionless_free_diffusion_exit_rate}\\
\aexitDL^{\trajvec,(0)} & =\evone^{(2)}
+\frac{\langle\efone,\efonedot\rangle^{(0)}}{\langle\efone,\efone\rangle^{(0)}}
- \frac{\dot{\IntDL}_{1}^{(0)}}{\IntDL_{1}^{(0)}}
\\
\aexitDL^{\trajvec,(2)} & =\evone^{(4)}+
\label{eq:dimensionless_N_dimension_exit_rate_order_two}
\frac{\langle\efone,\efonedot\rangle^{(2)}}{\langle\efone,\efone\rangle^{(0)}}
-\frac{\langle\efone,\efonedot\rangle^{(0)}}{\langle\efone,\efone\rangle^{(0)}}
\frac{\langle\efone,\efone\rangle^{(2)}}{\langle\efone,\efone\rangle^{(0)}}
\\ &\quad  \nonumber
-\frac{\dot{\IntDL}_{1}^{(2)} - \dot{\SumDL}^{(0)}}{\IntDL_{1}^{(0)}}
-\frac{{\IntDL}_{1}^{(2)} - {\SumDL}^{(0)}}{\IntDL_{1}^{(0)}}\left( 2 \frac{\dot{\epsilonDL}}{\epsilonDL}
	 -\frac{\dot{\IntDL}_{1}^{(0)}}{\IntDL_{1}^{(0)}}
\right).
\end{align}
Here, we define
\begin{align}
\IntDL_{n}(\tDL) &\equiv\int_{\BDL}\mathrm{d}^{N}\xvecDL~\efn(\xDLvec,\tDL),\label{eq:definition_integral_n}
\\
\SumDL(\tDL) &\equiv \sum_{n=2}^{\infty} 
\label{eq:def_sum_coupling}
 \left. \frac{1}{\Lnone (\tDL)} \frac{\langle \efn,\efonedot\rangle}{\langle \efn,\efn\rangle}\right|_{\tDL}\, 
 \IntDL_n(\tDL),
\end{align}
and by a superscript $(k)$ we 
denote the $k$-th order term in an expansion in powers of $\epsilonDL$, e.g.
$\IntDL_{n}^{(k)}(\tDL)
\equiv \int_{\BDL}\mathrm{d}^{N}\xvecDL~\efnk(\xDLvec,\tDL)$
(the perturbative eigenfunction at order $k$, $\efnk$, is
defined in Eq.~\eqref{eq:evn_power_series} below).
To obtain the power series expansion Eqs.~(\ref{eq:dimensionless_N_dimesional_exit_rate_power_series}-\ref{eq:dimensionless_N_dimension_exit_rate_order_two}) we
have used the parity properties of the 
perturbative spectrum; these are derived using the identical strategy
 as employed for the same purpose
 in Ref.~\cite{kappler_stochastic_2020}.

With Eqs.~(\ref{eq:dimensionless_N_dimesional_exit_rate_power_series}-\ref{eq:dimensionless_N_dimension_exit_rate_order_two}), we have an expression of the exit rate
 in terms of the instantaneous perturbative FP spectrum;
the exit rate in physical units is obtained 
using Eq.~\eqref{eq:exit_rate_DL} and the definitions of the dimensionless quantities given at the beginning
of the present appendix.
If the diffusion tensor is isotropic and independent of the state, $\Dmat \equiv D_0 \underline{\mathbb{1}}$,
and if the tube radius is constant, $\dot{R} = 0$,
then the exit rate given here reduces to the previous result from
Ref.~\cite{kappler_stochastic_2020}.

In App.~\ref{app:N_dim_spectrum} we derive the equations 
which determine the 
instantaneous 
spectrum
$(-\evn,\efn)$
perturbatively as a power series in $\epsilonDL$.
In particular, we show that
at time $\tDL = t/\td$, Eq.~\eqref{eq:dimensionless_free_diffusion_exit_rate}
is the steady-state free-diffusion exit rate from a ball of radius $R(t)$ and with 
a diffusion tensor given by $\Dmat(\trajvec(t),t)$.
Equation \eqref{eq:dimensionless_N_dimesional_exit_rate_power_series}
thus shows explicitly 
that this instantaneous steady-state free-diffusion exit rate
dominates the exit rate
 for small enough
tube radius.

\section{Perturbative FP spectrum and
reflecting-boundary steady state}
\label{app:N_dim_spectrum}

\subsection{Perturbative FP spectrum.}
To derive a perturbation series for the instantaneous spectrum
 Eq.~\eqref{eq:dimensionless_FPO_eigenvalue_equation}, 
we now generalize the derivation
from Ref.~\cite{kappler_stochastic_2020}, 
which considered diffusive dynamics with additive isotropic noise.
For this, we perform a spatial Taylor expansion
of both the apparent drift and the diffusion tensor
around the reference path $\trajvec$ in the eigenvalue
Eq.~\eqref{eq:dimensionless_FPO_eigenvalue_equation},
then substitute power series expansions for both the eigenvalue and eigenvector,
\begin{align}
\label{eq:evn_power_series}
	\evn &= \sum_{k=0}^{\infty} \epsilonDL^k \evn^{(k)},\qquad
	\efn = \sum_{k=0}^{\infty} \epsilonDL^k \efnk,
\end{align}
and demand that the resulting equation hold at each power of $\epsilonDL$
separately.
This yields a hierarchy of equations which at order $k$ is given  by
\begin{align}
 \label{eq:spectrum_hierarchy}
&\Dexp^{(0)}_{ij}\partialDL_i\partialDL_j\efnk+\evn^{(0)}\efnk  =-\sum_{l=1}^{k}\evn^{(l)}\efn^{(k-l)}  \\
 & -\sum_{l=1}^{k}l\,\EDL^{i}_{l,\alphavec} \partialDL_i\left(\xDL_{\alphavec}\efn^{(k-l)}\right)
\nonumber
- 
 \sum_{\substack{m \geq0\\
 l \geq 1\\
l+m=k
}
}
\Dexp^{(l)}_{ij,\alphavec}\,
\partialDL_i\partialDL_j \left[ \xDL_{\alphavec} \efn^{(m)}\right],
\end{align}
where the sums on the right-hand side are zero for $k=0$, we define
\begin{align}
\EDL^{i}_{k,\alphavec}(\tDL)  &\equiv \EDL^{i}_{k,\alpha_{1} \cdots\,\alpha_{k-1}}(\tDL)
\\
&\equiv -\left. \frac{L^{k-1} \td}{k!}
\nonumber
 \frac{\partial^{k-1}a_i}{\partial x_{\alpha_{1}}\cdots \partial x_{\alpha_{k-1}}}\right|_{(\bm{\traj}(t),t)}
\\
&
\label{eq:definition_EDL_N_dimensional}
\quad
+\delta_{k,1}\trajdotDL_{i}(\tDL)
 + \frac{\delta_{k,2}\delta_{i,\alpha_1}}{2}\frac{ \dot{\epsilonDL}(\tDL)}{\epsilonDL(\tDL)},
 \\
 \Dexp^{(k)}_{ij,\alphavec}(\tDL)
&\equiv 
	\Dexp^{(k)}_{ij,\alpha_1\cdots \alpha_{k}}(\tDL) 	
	 \label{eq:taylor_expansion_multidim_D_coeffs_app} 
	\equiv
		\frac{L^k}{k! \,D_0}\left.\frac{ \partial^k \,D_{ij}}{\partial x_{\alpha_1} \cdots \partial x_{\alpha_k}}\right|_{(\trajvec(t),t)},
\end{align}
and where we use the Einstein sum convention for the indices $\alphavec$.
At this point it becomes relevant that we assume $R(t) = R_0 r(t)$,
which implies that 
$\dot{\epsilonDL}/\epsilonDL \equiv \td\dot{r}/r$ 
in Eq.~\eqref{eq:definition_EDL_N_dimensional}
is independent of the perturbation parameter $R_0$.

The eigenfunction contribution at order $k$ needs
to fulfill the absorbing boundary conditions
$\efnk(\xvecDL,\tDL)=0$
for $||\xvecDL||_2=1$.
To close the system of equations 
defined by Eq.~\eqref{eq:spectrum_hierarchy} and the absorbing boundary conditions,
we introduce the normalization condition $\langle \efn,\efn\rangle = 1$,
which can be expanded as a power series in $\epsilonDL$ to yield a condition
for each $\efnk$ \cite{kappler_stochastic_2020}.
As we discuss in the remainder of this appendix, 
the spectrum is then calculated to arbitrary order by
solving this system of equations recursively.

At the lowest order, $k=0$, Eq.~\eqref{eq:spectrum_hierarchy}
 reduces to the eigenvalue equation
of the anisotropic Laplace operator,
\begin{align}
\label{eq:zeroth_order_spectrum_eq_app}
\Dexp^{(0)}_{ij}&\partialDL_i\partialDL_j\efn^{(0)}= -\evn^{(0)}\efn^{(0)}.
\end{align}
 Thus, $\evn^{(0)}$, $\efn^{(0)}$ is the
spectrum of the anisotropic Laplace operator in a unit ball with absorbing boundary conditions.
From Eq.~\eqref{eq:zeroth_order_spectrum_eq_app} we see that,  at time $\tDL$, to lowest order the
spectrum is that of free diffusion with a diffusion tensor $\Dexp^{(0)}_{ij}(\tDL) \equiv D_{ij}(\trajvec(t),t)/D_0$;
 in particular, $\evone^{(0)}$ is the corresponding instantaneous steady-state free-diffusion exit rate.
 Using Eq.~\eqref{eq:dimensionless_free_diffusion_exit_rate}, the definition of the function $f$ in
Eq.~\eqref{eq:afree_intro} thus follows.
Because the diffusion matrix $D_{ij}$
is by definition symmetric, it can be diagonalized
via an eigenbasis that is orthonormal with 
respect to the standard Euclidean inner product.
By expressing
Eq.~\eqref{eq:zeroth_order_spectrum_eq_app}
with respect to such an eigenbasis of $D_{ij}$, 
and subsequently
rescaling each axis by the corresponding eigenvalue (which for a full-rank
diffusivity tensor is positive),
 it follows that the equation is equivalent to
the eigenvalue equation for the Laplace operator
in an $N$-dimensional ellipsoid (with absorbing boundary conditions).

Once the spectrum has been calculated to order $\epsilonDL^{k-1}$,
the subsequent order is obtained in two steps \cite{kappler_stochastic_2020}.
First, we obtain an equation for the eigenvalue contribution $\evn^{(k)}$ 
by multiplying Eq.~\eqref{eq:spectrum_hierarchy} with $\efn^{(0)}$ and subsequently
 integrating over $\xvecDL$.
This yields 
\begin{align}
\evn^{(k)} & =-\sum_{l=1}^{k-1}\evn^{(l)}\int_{\BDL}\mathrm{d}^N\xvecDL~\efn^{(0)}\efn^{(k-l)}\\
 & \qquad-\sum_{l=1}^{k}l\,\int_{\BDL}\mathrm{d}^{N}\xvecDL~\efn^{(0)}\EDL^i_{l,\alphavec} \partialDL_i\left(\xDL_{\alphavec}\efn^{(k-l)}\right) 
 \nonumber 
  \\ &\qquad \nonumber
-
 \sum_{\substack{m \geq0\\
 l \geq 1\\
l+m=k
}
}
\Dexp^{(l)}_{ij,\alphavec}\,
\int_{\BDL}\mathrm{d}^N\xvecDL~\left[
\efn^{(0)}
\partialDL_i\partialDL_j \left( \xDL_{\alphavec}\efn^{(m)}\right)
\right],
\end{align}
which for constant isotropic diffusion reduces to the corresponding
result in Ref.~\cite{kappler_stochastic_2020}.
Because on the right-hand side of the equation, only the spectrum up to 
order $\epsilonDL^{k-1}$ appears, this equation can readily be used
to calculate $\evn^{(k)}$.
The result is then substituted in Eq.~\eqref{eq:spectrum_hierarchy}, and 
$\efnk$ is the solution to the
 resulting
inhomogeneous anisotropic Helmholtz equation 
with absorbing boundary conditions.

\subsection{Reflecting-boundary steady state.}
To evaluate the inner product 
Eq.~\eqref{eq:inner_product_definition}
perturbatively, the expansion of $\reqDL^{-1} \equiv 1 /\reqDL$
in powers of $\epsilonDL$ needs to be known.
The instantaneous reflecting-boundary steady state of the
dimensionless FP operator Eq.~\eqref{eq:dimensionless_Ndim_FPO} is defined by
\begin{align}
\partialDL_i \tilde{\text{J}}_{\mathrm{ss},i} 
 =0,\label{eq:req_defining_equation_appendix}
\end{align}
with
$\tilde{\text{J}}_{\mathrm{ss},i} = 
\epsilonDL\,\aDL_{\mathrm{app},i} \,\reqDL
-\partialDL_j ( \DDL_{ij} \reqDL)$.
The corresponding reflecting boundary conditions are
\begin{equation}
\left.\left[\hat{{n}}_i \tilde{\text{J}}_{\mathrm{ss},i}\right] \right|_{\partial\BDL}=0,\label{eq:req_boundary_condition_appendix} 
\end{equation}
with $\hat{n}_i$ the $i$-th component of the
 outward-pointing unit normal vector on the unit sphere $\partial \BDL$.
Similar to the perturbative calculation of the spectrum, a hierarchy 
of equations for the coefficients $\reqDL^{(k)}$ of the series expansion
of $\reqDL$ in powers of $\epsilonDL$ 
is obtained from Eqs.~\eqref{eq:req_defining_equation_appendix}, \eqref{eq:req_boundary_condition_appendix}
by substituting expansions in powers of $\epsilonDL$, and demanding the resulting equation hold
at each power of $\epsilonDL$ separately.
At order $k$, we obtain
\begin{align}
\Dexp^{(0)}_{ij} &\partialDL_i \partialDL_j  \reqDL^{(k)}
=
-\sum_{l=1}^{k}l\,\EDL^i_{l,\alphavec}\,\partialDL_i \left(\xDL_{\alphavec}\reqDL^{(k-l)}\right)\label{eq:req_hierarchy_appendix}
\\ & \nonumber \qquad\qquad \qquad
-
 \sum_{\substack{m \geq0\\
 l \geq 1\\
l+m=k
}
}
\Dexp^{(l)}_{ij,\alphavec}\,
\partialDL_i\partialDL_j \left[ \xDL_{\alphavec}  \efn^{(m)}\right],
\end{align}
where the right-hand side is zero for $k = 0$.
This equation defines $\reqDL^{(k)}$, the corresponding boundary conditions 
follow from Eq.~\eqref{eq:req_boundary_condition_appendix} as
\begin{align}
0 =&
-\left.\sum_{l=1}^{k}l\,\EDL^i_{l,\alphavec}\,\left[ \hat{n}_i \left(\xDL_{\alphavec}\reqDL^{(k-l)}\right)\right]\right|_{\xDL }\label{eq:req_hierarchy_bc_appendix}
\\ & \nonumber
-
 \sum_{\substack{l, m \geq0\\
l+m=k
}
}
\Dexp^{(l)}_{ij,\alphavec}\,
\left.
\vphantom{\sum_{l=1}^{k}}
\left[
\hat{n}_i\partialDL_j \left( \xDL_{\alphavec}  \efn^{(m)}\right)\right]\right|_{\xDL},
\end{align}
where $\xDL \in \partial \BDL$ and
where the first sum on the right-hand side of the equation is zero for $k=0$.
Starting from the unnormalized solution $\reqDL^{(0)} \equiv 1$ at order $\epsilonDL^0$,
this system of equations can be solved recursively.
From the resulting perturbation series for $\reqDL$, a perturbation series for
$\reqDL^{-1}$ is then obtained via the definition of the inverse, $\reqDL^{-1} \equiv 1/\reqDL$.
For additive isotropic noise, where 
$\Dexp^{(k)}_{ij} \equiv \delta_{i,j}\delta_{k,0}$, 
Eqs.~\eqref{eq:req_hierarchy_appendix}, \eqref{eq:req_hierarchy_bc_appendix} simplify to 
their counterparts in
Ref.~\cite{kappler_stochastic_2020}.

\section{Results for one-dimensional systems}
\label{app:one_dim}

In this appendix, we consider the special case of a one-dimensional system, $N=1$.
We
 derive explicit formulas for the perturbative FP spectrum and the exit rate.
We provide a python module named PyTubular, which contains symbolic implementations 
of the analytical results from this appendix \cite{module_pytubular}. 
Beyond the results derived in this appendix, the module PyTubular also contains the normalized version
of the 
perturbative propagator Eq.~\eqref{eq:perturbative_Ndim_propagator}.

\subsection{Perturbative spectrum of the 1D FP equation.}
For a one-dimensional system, $N=1$,
we solve Eq.~\eqref{eq:spectrum_hierarchy} 
recursively using
 the same algorithm as employed in Ref.~\cite{kappler_stochastic_2020} 
 for the simplified case of additive noise.
 We now give the resulting lowest order contributions to the both eigenvalues and eigenfunctions; 
the analytical spectrum up to including order $\epsilonDL^5$ is available in the python module 
PyTubular \cite{module_pytubular}.

From the parity properties of the spectrum, which are derived
as in Ref.~\cite{kappler_stochastic_2020},
it follows that $\evn^{(k)} = 0$ for odd $k$.
For even $k$, the first two nonzero contributions to the eigenvalue are
\begin{align}
\label{eq:evn0_app}
	\evn^{(0)} &= \Dexp_0 \left( \frac{ n \pi}{2}\right)^2,
\end{align}
\begin{align}
	\label{eq:evn2_app}
	\Dexp_0\evn^{(2)} &=
	 \frac{\EDL_1^2}{4}
					- \Dexp_0\EDL_2
					+\frac{1}{16} \left(  - \pi^2 n^2 + 3 \right)\Dexp_1^2
					\\& \quad \nonumber
					+ \frac{1}{12}\left( \pi^2 n^2 - 6 \right)\Dexp_0 \Dexp_2
					+ \frac{1}{2}\Dexp_1 \EDL_1,
\end{align}
where according to Eqs.~\eqref{eq:definition_EDL_N_dimensional}, \eqref{eq:taylor_expansion_multidim_D_coeffs_app} 
we have
\begin{align}
\EDL_{k}(\tDL) &\equiv-\left. \frac{L^{k}}{D_0 k!} \frac{\partial^{k-1}a}{\partial x^{k-1}}\right|_{(\traj(t),t)}
\label{eq:definition_EDL_N_dimensional_1D} 
\\
& \qquad\nonumber
+\delta_{k,1}\frac{\td}{L}\dot{\traj}(t)
 + \td\frac{\delta_{k,2}}{2}\frac{ \dot{r}(t)}{r(t)},
 \\
\label{eq:Dexp_def_app}
\Dexp_k(\tDL) &\equiv \Dexp^{(k)}(\tDL) \equiv \frac{1}{k!}\frac{L^k}{ D_0} \left.\frac{\partial^{k}D}{\partial x^{k}}\right|_{(\traj(t),t)}.
\end{align}

As in the simpler case of additive noise \cite{kappler_stochastic_2020}, 
the $k$-th order term of $\efn$ is of the form
\begin{align}
\efnk(\xDL,\tDL) & =\Qsk(\xDL,\tDL)\,\sin\left[n\frac{\pi}{2}(\xDL+1)\right]\\
 \nonumber& \quad+\Qck(\xDL,\tDL)\,\cos\left[n\frac{\pi}{2}(\xDL+1)\right], 
\end{align}
where the prefactors $\Qsk$, $\Qck$ are polynomials
in $\xDL$.
Up to order $\epsilonDL^2$, they are given by
\begin{align}
\label{eq:Qc0}
\Qs^{(0)}(\xDL) &= 1, \qquad  \qquad
\Qc^{(0)}(\xDL) = 0,\\
\label{eq:Qs1}
\Qs^{(1)}(\xDL) &= -\frac{\xDL^2}{4 \Dexp_0}\left(
 2 \EDL_1
+ 3 \Dexp_1
\right),
\\
\label{eq:Qc1}
\Qc^{(1)}(\xDL) &= 
\frac{ n \pi \xDL}{8 \Dexp_0}
\left(1 - \xDL^2\right) \Dexp_1,
\end{align}
\begin{align}
\label{eq:Qs2}
\Qs^{(2)}&(\xDL) =
	\frac{\xDL}{384 \Dexp_0^2}
	\left[\vphantom{\frac{1}{2}} \right.
48 \xDL^2 \EDL_1^2
 + 240 \xDL^2 \Dexp_1 \EDL_1
 \\ &  \nonumber
 + \left( - 24 +252 \xDL^2 - 3 \pi^2 n^2 - 3 \pi^2 n^2 \xDL^4 +6 \pi^2 n^2 \xDL^2\right) \Dexp_1^2
\\ & \qquad \qquad \nonumber 
- 192 \xDL^2 \Dexp_0 \EDL_2
 + \left(32 - 288 \xDL^2\right) \Dexp_0 \Dexp_2
 	\left.\vphantom{\frac{1}{2}}\right],
\end{align}
\begin{align}
\label{eq:Qc2}
\Qc^{(2)}&(\xDL) =
\frac{ \xDL^2 n \pi}{96 \Dexp_0^2}
\left( 1 -  \xDL^2\right)
	\left[\vphantom{\frac{1}{2}} \right.
-6 \Dexp_1 \EDL_1
 - 15\Dexp_1^2
 + 8  \Dexp_0 \Dexp_2
	\left.\vphantom{\frac{1}{2}}\right],
\end{align}
with $\EDL_k \equiv \EDL_k (\tDL)$, $\Dexp_k\equiv \Dexp_k(\tDL)$ given by
Eqs.~\eqref{eq:definition_EDL_N_dimensional_1D},  \eqref{eq:Dexp_def_app}.

If the diffusivity is independent of position and time, $D(x,t) \equiv D_0$,
then $\Dexp_0 \equiv 1$ and $\Dexp_k = 0$ for all $k \geq 1$.
In that case, Eqs.~(\ref{eq:evn0_app}-\ref{eq:Qc2}) 
 reduce to the corresponding results derived  in Ref.~\cite{kappler_stochastic_2020}.

\subsection{Reflecting-boundary steady state.}
For a one-dimensional system,
Eqs.~\eqref{eq:req_defining_equation_appendix}, \eqref{eq:req_boundary_condition_appendix}
are solved by
\begin{align}
\label{eq:reqDL_intermediate_app}
\DDL(\xDL) \reqDL(\xDL) &= 
\Dexp_0
\exp\left[ \int_0^{\xDL} \mathrm{d}\yDL\,\left.
\dfrac{ \epsilonDL  \aDL_{\mathrm{app}} \reqDL
}{\DDL}	\right|_{\yDL}
		\right].
\end{align}
The prefactor in Eq.~\eqref{eq:reqDL_intermediate_app} 
can be chosen arbitrarily; we choose it such that $\reqDL(\xDL = 0) = 1$,
a different choice corresponds to a rescaling of the inner product.
Solving Eq.~\eqref{eq:reqDL_intermediate_app} for $\reqDL^{-1}(\xDL)$ and
 substituting the power series 
 expansions for $\aDL_{\mathrm{app}}$, $\DDL$,
we obtain
\begin{align}
\label{eq:reqDL_intermediate3_app}
\reqDL^{-1}(\xDL) &= 
\frac{
\sum_{k=0}^{\infty} \epsilonDL^k \Dexp_k \xDL^{k}
}{
\Dexp_0
}
\exp\left[ \int_0^{\xDL} \mathrm{d}\yDL\,
\frac{ 
\sum_{k=1}^{\infty} k\epsilonDL^k 
 \EDL_k
\yDL^{k-1}
}{
\sum_{k=0}^{\infty} \epsilonDL^k \Dexp_k \yDL^{k}
}
\right].
\end{align}
From this, 
the power-series expansion 
of $\reqDL^{-1}$
is obtained by
first expanding the integrand in the exponent to the desired order in $\epsilonDL$, 
performing the integral in the exponent, 
subsequently expanding both the exponential and the prefactor,
 and finally expanding their product to the desired order in $\epsilonDL$.
To order $\epsilonDL^2$, this yields
\begin{align}
\left(\reqDL^{-1}\right)^{(0)} &= 1, \qquad
\left(\reqDL^{-1}\right)^{(1)} = \frac{\xDL}{\Dexp_0}\left(
\EDL_1
 + \Dexp_1\right),
\\
\left(\reqDL^{-1}\right)^{(2)} &= 
\frac{\xDL^2}{2 \Dexp_0^2}
\left(
 \EDL_1^2
 + \Dexp_1 \EDL_1
 +2 \Dexp_0 \EDL_2
 +2 \Dexp_0 \Dexp_2 
\right)
\end{align}
with $\EDL_k \equiv \EDL_k (\tDL)$, $\Dexp_k\equiv \Dexp_k(\tDL)$ given by
Eqs.~\eqref{eq:definition_EDL_N_dimensional_1D},  \eqref{eq:Dexp_def_app}.

\subsection{Exit rate to order $R^2$ for one-dimensional systems.}
Employing the perturbative results for the one-dimensional spectrum and 
instantaneous steady state discussed just above,
the exit rate Eqs.~(\ref{eq:dimensionless_N_dimesional_exit_rate_power_series}-\ref{eq:dimensionless_N_dimension_exit_rate_order_two}) is readily evaluated.
The resulting terms of the perturbation series are
\begin{align}
\label{eq:afree_DL_app}
\tilde{\alpha}_{\mathrm{free}}^{\traj} 
& =
\frac{\pi^2 \Dexp_0}{4 \epsilonDL^2},
\end{align}
\begin{widetext}
\begin{align}
\label{eq:alpha0_app}  
	\Dexp_0 \,\tilde{\alpha}^{\traj,(0)} &=
	 \frac{\EDL_1^2}{4}
	- \Dexp_0 \EDL_2
	+ \frac{\Dexp_1 \EDL_1}{2}
	- \left( \frac{\pi^2}{16} - \frac{3}{16}\right) \Dexp_1^2
	+ \left( \frac{\pi^2}{12} - \frac{1}{2}\right) \Dexp_0\Dexp_2,
\\
\label{eq:alpha2_app}
	\Dexp_0^3 \,\tilde{\alpha}^{\traj,(2)} &=
\left(\frac{1}{8}  -\frac{ 3}{8 \pi^2} \right) \Dexp_1^2 \EDL_1^2
 + \left(\frac{1}{4}  -\frac{ 3}{4 \pi^2} \right) \Dexp_1^3 \EDL_1
 + \left(\frac{3}{32}  -\frac{ 9}{32 \pi^2}  -\frac{ \pi^2}{64} \right) \Dexp_1^4
 + \left( -\frac{ 1}{12}  +\frac{1}{2 \pi^2} \right) \Dexp_0 \Dexp_2 \EDL_1^2
\\ & \quad \nonumber 
 + \left( -\frac{ 1}{2}  +\frac{3}{2 \pi^2} \right) \Dexp_0 \Dexp_1 \EDL_1 \EDL_2
 + \left( -\frac{ 2}{3}  +\frac{5}{2 \pi^2} \right) \Dexp_0 \Dexp_1 \Dexp_2 \EDL_1
 + \left( -\frac{ 1}{2}  +\frac{3}{2 \pi^2} \right) \Dexp_0 \Dexp_1^2 \EDL_2
\\ & \quad \nonumber 
 + \left( -\frac{ 7}{16}  +\frac{\pi^2}{16}  +\frac{3}{2 \pi^2} \right) \Dexp_0 \Dexp_1^2 \Dexp_2
 + \left(\frac{1}{3}  -\frac{ 2}{\pi^2} \right) \Dexp_0^2 \EDL_2^2
 + \left(\frac{2}{3}  -\frac{ 4}{\pi^2} \right) \Dexp_0^2 \Dexp_2 \EDL_2
 + \left(\frac{1}{4}  -\frac{ 3}{2 \pi^2}  -\frac{ \pi^2}{60} \right) \Dexp_0^2 \Dexp_2^2
\\ & \quad \nonumber 
 + \left(\frac{1}{2}  -\frac{ 3}{\pi^2} \right) \Dexp_0^2 \EDL_1 \EDL_3
 + \left(\frac{1}{2}  -\frac{ 3}{\pi^2} \right) \Dexp_0^2 \Dexp_3 \EDL_1
 + \left(1 -\frac{ 3}{2 \pi^2} \right) \Dexp_0^2 \Dexp_1 \EDL_3
 + \left(\frac{5}{8}  -\frac{ 3}{2 \pi^2}  -\frac{ 3 \pi^2}{40} \right) \Dexp_0^2 \Dexp_1 \Dexp_3
\\ & \quad \nonumber 
 + \left( - 2 +\frac{12}{\pi^2} \right) \Dexp_0^3 \EDL_4
 + \left( - 1 +\frac{6}{\pi^2}  +\frac{\pi^2}{20} \right) \Dexp_0^3 \Dexp_4
 + \left(\frac{1}{3}  -\frac{ 3}{\pi^2} \right) \Dexp_0^2 \dot{\EDL}_2
 + \left(\frac{1}{6}  -\frac{ 1}{\pi^2} \right) \Dexp_0^2 \dot{\Dexp}_2
\\ & \quad \nonumber 
 + \left( -\frac{ 1}{4}  +\frac{2}{\pi^2} \right) \Dexp_0 \EDL_1 \dot{\EDL}_1
 + \left( -\frac{ 5}{24}  +\frac{7}{4 \pi^2} \right) \Dexp_0 \Dexp_1 \dot{\EDL}_1
 + \left( -\frac{ 7}{24}  +\frac{3}{2 \pi^2} \right) \Dexp_0 \EDL_1 \dot{\Dexp}_1
 + \left( -\frac{ 3}{16}  +\frac{3}{4 \pi^2} \right) \Dexp_0 \Dexp_1 \dot{\Dexp}_1
\\ & \quad \nonumber 
 + \left(\frac{1}{4}  -\frac{ 2}{\pi^2} \right) \EDL_1^2 \dot{\Dexp}_0
 + \left(\frac{1}{2}  -\frac{ 13}{4 \pi^2} \right) \Dexp_1 \EDL_1 \dot{\Dexp}_0
 + \left(\frac{3}{16}  -\frac{ 3}{4 \pi^2} \right) \Dexp_1^2 \dot{\Dexp}_0
 + \left( -\frac{ 1}{3}  +\frac{3}{\pi^2} \right) \Dexp_0 \EDL_2 \dot{\Dexp}_0
\\ & \quad \nonumber 
 + \left( -\frac{ 1}{6}  +\frac{1}{\pi^2} \right) \Dexp_0 \Dexp_2 \dot{\Dexp}_0
 + \left( -\frac{ 1}{4}  +\frac{2}{\pi^2} \right) \Dexp_0 \EDL_1^2 \frac{ \dot{\epsilonDL}}{\epsilonDL}
 + \left( -\frac{ 3}{4}  +\frac{4}{\pi^2} \right) \Dexp_0 \Dexp_1 \EDL_1 \frac{ \dot{\epsilonDL}}{\epsilonDL}
 + \left( -\frac{ 7}{16}  +\frac{3}{2 \pi^2} \right) \Dexp_0 \Dexp_1^2 \frac{ \dot{\epsilonDL}}{\epsilonDL}
\\ & \quad \nonumber 
 + \left(1 -\frac{ 8}{\pi^2} \right) \Dexp_0^2 \EDL_2 \frac{ \dot{\epsilonDL}}{\epsilonDL}
 + \left(\frac{2}{3}  -\frac{ 4}{\pi^2} \right) \Dexp_0^2 \Dexp_2 \frac{ \dot{\epsilonDL}}{\epsilonDL},
\end{align}
\end{widetext}
with $\EDL_k \equiv \EDL_k (\tDL)$, $\Dexp_k\equiv \Dexp_k(\tDL)$ given by
Eqs.~\eqref{eq:definition_EDL_N_dimensional_1D},  \eqref{eq:Dexp_def_app},
and as before a dot on a dimensionless quantity denotes a derivative with respect to
dimensionless time $\tDL$.
By substituting the definitions of $\EDL_k$, $\Dexp_k$, the exit rate is fully
expressed in terms of the drift and diffusion of the 1D FPE.
In particular, from Eq.~\eqref{eq:afree_DL_app} we obtain the 
steady-state free diffusion exit rate
\begin{align}
\alpha_{\mathrm{free}}^{\traj} &=
\frac{1}{\td} \tilde{\alpha}_{\mathrm{free}}^{\traj} 
=\frac{\pi^2}{4} \frac{D(\traj(t),t)}{R(t)^2},
\end{align}
so that the function $f$ from Eq.~\eqref{eq:afree_intro} is for $N=1$ given by
$f(M) = \pi^2 M/4$ and Eq.~\eqref{eq:afree_1D} follows.
The order-$R^0$ contribution Eq.~\eqref{eq:Lagrangian_1D} is obtained by substituting the
definitions of $\EDL_k$, $\Dexp_k$ into Eq.~\eqref{eq:alpha0_app} and using the 
 drift and diffusion of the FP Eq.~\eqref{eq:1D_FP_eq}.

\subsection{Comparison of small-radius perturbative results to numerical simulations.}
In Sect.~\ref{sec:1D_exit_rate} we consider the exit rate Eq.~\eqref{eq:axit_result_intro}
to order $R^0$ for three different scenarios,
namely a constant-radius tube, a constant free-diffusion exit rate tube,
and the exit rate related to the Stratonovich construction.
We now compare 
our perturbative analytical results to order $R^0$
with exit rates measured in numerical 
simulations.
The results demonstrate
 that for all three scenarios,
and for the radii considered in Sect.~\ref{sec:1D_exit_rate},
 the analytical perturbative exit rate Eq.~\eqref{eq:axit_result_intro} to order $R^0$ 
  describes the actual exit rate very well.

For the numerical results presented here, the dimensionless form of the
 one-dimensional FP Eq.~\eqref{eq:1D_FP_eq} with absorbing boundary conditions
 is simulated using the forward Euler algorithm 
described in Ref.~\cite{kappler_stochastic_2020};
from the resulting trajectory, the exit rate is obtained via numerical evaluation 
 of Eq.~\eqref{eq:inst_exit_rate}.

\begin{figure*}[ht]
\centering \includegraphics[width=1\textwidth]{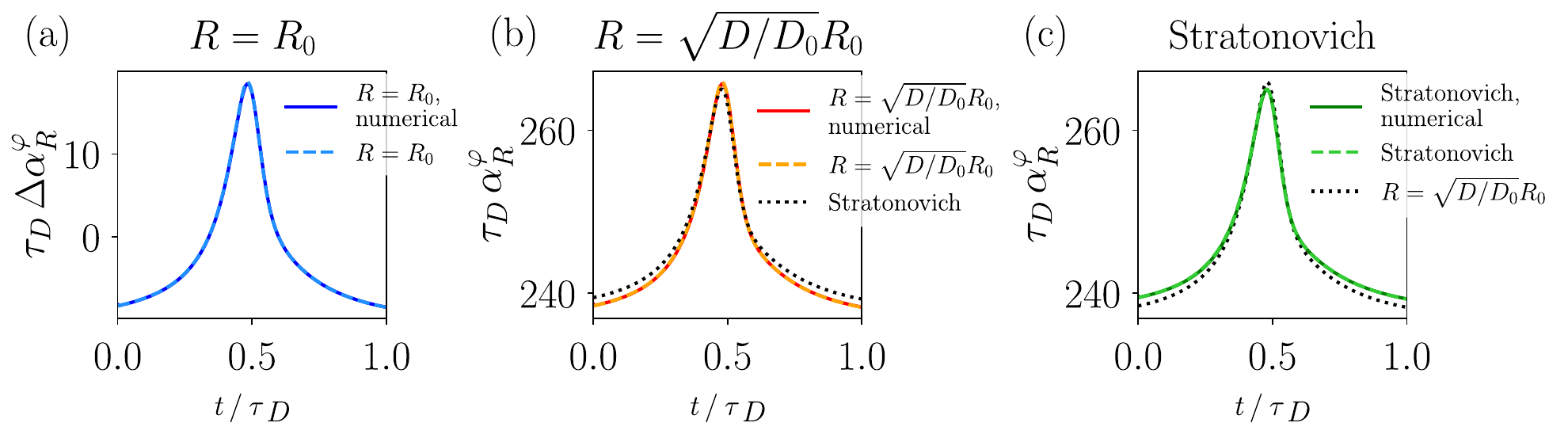} 
\caption{ \label{fig:exit_rate_numerical} 
Comparison of theoretical and numerical exit rates for the example system
considered in Sect.~\ref{sec:numerical_example}.
All subplots compare the 
exit rates
measured in numerical simulations (solid lines)
to the corresponding theoretical prediction $\alpha^{\traj,(0)}$ (dashed lines).
For numerical results, the dimensionless 
 FP Eq.~\eqref{eq:1D_FP_eq} with absorbing boundary conditions
 is simulated using the algorithm 
described in Ref.~\cite{kappler_stochastic_2020};
from the resulting trajectory, the exit rate is obtained via numerical evaluation 
 of Eq.~\eqref{eq:inst_exit_rate}.
Subplot (a) shows results for constant tube radius (scenario 1).
The solid line depicts the theoretical result Eq.~\eqref{eq:scenario1_Lagrangian},
the dashed line is obtained by measuring the exit rate in numerical simulations
and subtracting Eq.~\eqref{eq:scenario1_free_diffusion}.
Subplot (b) considers a time-dependent tube radius such that the
free-diffusion exit rate is constant (scenario 2).
The solid line depicts the sum of the theoretical results
 Eqs.~\eqref{eq:scenario2_free_diffusion}, \eqref{eq:scenario2_Lagrangian},
the dashed line is obtained by measuring the exit rate in numerical simulations.
For comparison, the Stratonovich exit rate Eq.~\eqref{eq:scenario3_aexit}  
is shown as dotted line.
Subplot (c) shows results for the Stratonovich construction
 (scenario 3).
The solid line depicts the Stratonovich exit rate Eq.~\eqref{eq:scenario3_aexit},
the dashed line is obtained by measuring the exit rate in numerical simulations.
For comparison, the exit rate obtained of the constant free-diffusion result 
Eqs.~\eqref{eq:scenario2_free_diffusion}, \eqref{eq:scenario2_Lagrangian}  
is shown as dotted line.
}
\end{figure*}

Scenario 1: Constant radius.
We now compare the analytical exit rate 
Eqs.~\eqref{eq:axit_result_intro}, \eqref{eq:scenario1_free_diffusion}, \eqref{eq:scenario1_Lagrangian},
to numerical results. 
In Fig.~\ref{fig:exit_rate_numerical} (a), we compare the perturbative result Eq.~\eqref{eq:scenario1_Lagrangian}
with
\begin{align}
\label{eq:exit_rate_subtract_free}
	\Delta \alpha_{R}^{\traj}(t) &= \alpha_R^{\traj}(t) - \aexitfree^{\traj}(t),
\end{align}
where $ \alpha_R^{\traj}$ is the numerical exit rate
and $\aexitfree^{\traj}$ is the analytical free-diffusion rate Eq.~\eqref{eq:scenario1_free_diffusion}.
According to the figure the theoretical 
expression $\mathcal{L}_1^{\traj,(0)}$
and the numerical result agree very well;
 for the radius $\epsilonDL \equiv R_0/L= 0.1$, 
the numerical exit rate is therefore fully described by the perturbative result 
Eqs.~\eqref{eq:axit_result_intro}, 
\eqref{eq:scenario1_free_diffusion},
\eqref{eq:scenario1_Lagrangian}.

Scenario 2: Constant free-diffusion exit rate.
In Fig.~\ref{fig:exit_rate_numerical} (b) we compare
the theoretical result Eq.~\eqref{eq:scenario2_Lagrangian}
with 
Eq.~\eqref{eq:exit_rate_subtract_free}, where 
$ \alpha_R^{\traj}$ is obtained from numerical simulation of scenario 2
and
$\aexitfree^{\traj}$ is given by Eq.~\eqref{eq:scenario2_free_diffusion}.
The excellent agreement between theoretical and numerical curves shows
that also in scenario 2, the perturbative expansion Eq.~\eqref{eq:axit_result_intro}
describes the actual exit rate from the small-radius tube.

Scenario 3: Stratonovich Lagrangian.
In Fig.~\ref{fig:exit_rate_numerical} (c) we compare the theoretical exit rate 
Eq.~\eqref{eq:scenario3_aexit} 
to results from numerical simulations of the FPE in the $y$-coordinate
defined in Eq.~\eqref{eq:coordinate_transformation_definition}.
As for scenarios 1 and 2, 
we 
find perfect agreement between perturbative
and numerical exit rates.
Thus, also for scenario 3 the perturbative expansion adequately describes
the actual exit rate for the small tube radius considered.

\subsection{MPT centers for order $R^0$ exit rate vs.~MPT centers for order $R^2$ exit rate.}
\begin{figure*}[ht]
\centering \includegraphics[width=1\textwidth]{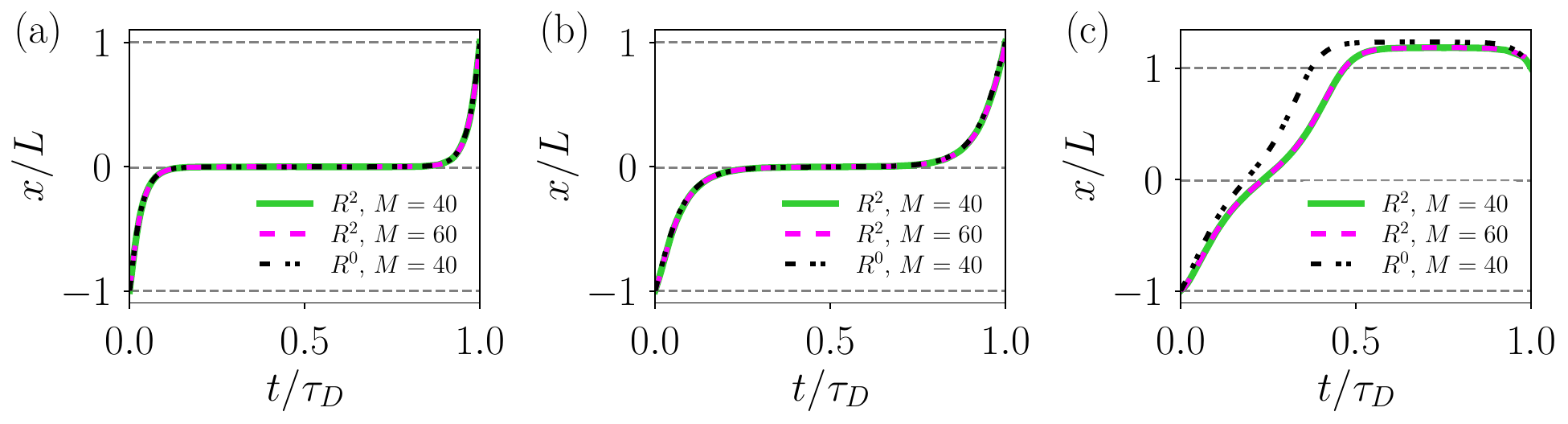} 
\caption{ \label{fig:instantons_R0_vs_R2} 
Comparison of MPT centers based on analytical theory for the example system
defined in Sect.~\ref{sec:numerical_example}.
All curves are obtained by minimizing the functional Eq.~\eqref{eq:Sdef_intro},
\eqref{eq:dimensionless_N_dimesional_exit_rate_power_series}
for constant radius (a) $R \equiv R_0 = 0.1 L$, 
(b) $R \equiv R_0 = 0.2L$, 
(c) $R  \equiv R_0 = 0.3L$
using the algorithm given in App.~\ref{app:functional_minimization_algorithm},
and paths that start at $\xl$ and end at $\xr$ after a duration $\td$.
The green solid lines are replots of the colored lines from 
Fig.~\ref{fig:radius_dependence_of_instanton},
 and are obtained via parametrizing path-space by $M = 40$ modes
and using the theoretical exit rate to order $R^2$.
While the magenta dashed lines use $M = 60$ modes
and the theoretical exit rate to order $R^2$,
the black dash-dotted lines use $M=40$ modes and the theoretical exit rate
to order $R^0$.
All terms in the perturbative exit rate 
Eq.~\eqref{eq:dimensionless_N_dimesional_exit_rate_power_series}
 are calculated via
\eqref{eq:afree_DL_app}, 
\eqref{eq:alpha0_app},
\eqref{eq:alpha2_app},
using the diffusivity and drift profiles
Eqs~\eqref{eq:D_example}, \eqref{eq:drift_potential}.
}
\end{figure*}

In Fig.~\ref{fig:radius_dependence_of_instanton} 
we discuss the radius-dependence of the MPT center
for constant-radius tubes,
based on the theoretical formula for the exit rate to order $R^2$ 
and a parametrization of path-space based on $M=40$ modes.
To assess the importance of both the order of the 
perturbative exit rate and the number of modes $M$ on the MPT center, we in
 Fig.~\ref{fig:instantons_R0_vs_R2} compare
 the results
 from Fig.~\ref{fig:radius_dependence_of_instanton}
with MPT centers based on 
(i)  the exit rate to order $R^2$ and using $M=60$ modes,
and (ii) the exit rate to order $R^0$ and using $M=40$ modes.
For radius $R \equiv R_0 = 0.1\,L$ and $R \equiv R_0 = 0.2\,L$,
we observe in Fig.~\ref{fig:instantons_R0_vs_R2} (a), (b)
that all three MPT centers agree perfectly with each other, which shows
that both $M= 40$ modes and the exit rate to order $R^0$ are
sufficient to describe MPT centers at these tube radii.
As we see in Fig.~\ref{fig:instantons_R0_vs_R2} (c), for $R = 0.3\,L$ 
the MPT centers based on the exit rate to order $R^2$ 
agree for both values $M = 40$, $M=60$,
which demonstrates that $M=40$ modes are sufficient to describe the most probable
tube center.
On the other hand, we see that the exit rate to order $R^0$ slightly disagrees with
the results based on the order $R^2$ theory; while the differences are not too big,
this shows that for this radius the quadratic term
Eq.~\eqref{eq:alpha2_app} in the perturbative series 
Eq.~\eqref{eq:dimensionless_N_dimesional_exit_rate_power_series} is already relevant.

\subsection{Transient initial decay of propagator}
\label{app:transient_decay}

The propagator Eq.~\eqref{eq:perturbative_Ndim_propagator}
is dominated by the dynamics of the slowest-decaying eigenmode
$\efone$ \cite{kappler_stochastic_2020}, and as such
is only valid after an initial relaxation time defined
 in Eq.~\eqref{eq:relaxation_time}.
The transient initial decay of the modes $n \geq 2$ is discussed in
the additive-noise derivation
of Eq.~\eqref{eq:perturbative_Ndim_propagator}
 in  Ref.~\cite{kappler_stochastic_2020};
since the discussion is based solely on the spectrum of the dimensionless FPE,
it carries over directly to the present case of multiplicative noise.
With this in mind, since for a one-dimensional system 
it holds that $\langle \efn, \efmdot \rangle = \mathcal{O}(\epsilonDL)$,
 the initial transient decay can also 
be included in the propagator.
This leads to
\begin{widetext}
\begin{align}
\label{eq:perturbative_one_dim_propagator}
	\PtubeDL
	&(\xvecDL,\tDL~\setbar~\xveczeroDL,\tzeroDL) =
	\frac{1}{\reqDL(\xzeroDLvec,\tinitialDL)}
\sum_{n=1}^{\infty}
	 \dfrac{	 \exp\left[ - \int_{\tinitialDL}^{\tDL}\mathrm{d}\tDLdummy~\dfrac{\Ln(\tDLdummy)}{\epsilonDL^2(\tDLdummy)} \right] 
}{ {\langle \efn,\efn \rangle|_{\tinitialDL}} } 
	 \\
	 & \times
 \left[ \efn(\xvecDL,\tDL) - \sum_{m=n+1}^{\infty}
    \nonumber
   \left.\frac{\epsilonDL^2(\tDL)}{\Lmn(\tDL)} 
   \frac{\langle \efm,\efndot\rangle}{\langle \efm,\efm\rangle}\right|_{\tDL}
   \left[
   1 - \exp\left( -
      	\frac{ \Lmn(\tDL) }{\epsilonDL^2(\tDL)} 
    \left( \tDL - \tinitialDL \right) 
	\right)
	   \right]
   \efm(\xvecDL,\tDL) \right]
   \\ & \times
 \left[ \efn(\xveczeroDL,\tinitialDL) -\sum_{m=n+1}^{\infty}
 \nonumber
   \left.
   \frac{\epsilonDL^2(\tinitialDL) }{\Lnm(\tinitialDL)}
   \frac{\langle \efn,\efmdot\rangle}{\langle \efm,\efm\rangle}\right|_{\tinitialDL}
   \left[
   1 - \exp\left( -
      	\frac{ \Lmn(\tinitialDL) }{\epsilonDL^2(\tinitialDL)} 
    \left( \tDL - \tinitialDL \right) 
	\right)
	   \right]
   \efm(\xveczeroDL,\tinitialDL) \right] 
   \\
	 \nonumber
   	    &+ \mathcal{O}(\epsilonDL_0^6).
\end{align}
\end{widetext}
In deriving Eq.~\eqref{eq:perturbative_one_dim_propagator}, we treat terms proportional
 to $(\tDL - \tinitialDL)^k \exp\left[ -( \tDL - \tinitialDL)
      	 \kappa/\epsilonDL_0^2\right]$
	(for some constant $\kappa = \mathcal{O}(\epsilonDL^0)$)
 as order $\epsilonDL_0^{2k}$, because 
	such terms are only non-negligible for $\tDL - \tinitialDL \lesssim \epsilonDL_0^2/\kappa 
	= \mathcal{O}(\epsilonDL_0^2)$.

Since $\Delta \tilde{\Lambda}_{21} = \min_{n \neq m} \Lmn$
and $\Delta \tilde{\Lambda}_{21} < \tilde{\Lambda}_l$ for all $l > 1$ and small
$\epsilonDL_0$,
Eq.~\eqref{eq:perturbative_one_dim_propagator} reduces
 to Eq.~\eqref{eq:perturbative_Ndim_propagator} in the limit 
Eq.~\eqref{eq:relaxation_time}.

\section{Algorithm for functional minimization in path space}
\label{app:functional_minimization_algorithm}

In Sect.~\ref{sec:most_probable_tube} we consider the most 
probable tube center as defined in Eq.~\eqref{eq:def_most_probable_tube},
with the action given by Eq.~\eqref{eq:Sdef_intro} as
the temporal integral over the tubular exit rate.
For a one-dimensional system the first three terms in the perturbative 
tubular exit rate Eq.~\eqref{eq:dimensionless_N_dimesional_exit_rate_power_series}
are given in terms of the diffusivity and drift 
by 
Eqs.~\eqref{eq:afree_DL_app}, \eqref{eq:alpha0_app},
\eqref{eq:alpha2_app}.

While for given diffusivity, drift, and path it is thus straightforward to 
evaluate the action, it is not straightforward to perform the functional minimization
Eq.~\eqref{eq:def_most_probable_tube}
which is over the infinite-dimensional space of continuous paths 
 that start 
at $\xl$ and end at $\xr$ after duration $\tfinal$.
We approximate this infinite-dimensional space
 by considering paths parametrized via \cite{kappler_measurement_2022}
\begin{equation}
\traj(t) = \xl + \frac{ t}{\tfinal} (\xr - \xl) 
+ 
\sum_{n=1}^{M} \frac{c_n}{n^2} \sin\left( n \pi  t/\tfinal \right),
\end{equation}
so that a path is represented by a finite-dimensional vector of mode coefficients
 $(c_1, c_2, ..., c_M) \in \mathbb{R}^M$. 
 Upon substituting this path-space parametrization
 into any of the analytical formulae for the exit rate,
 finding the minimum in Eq.~\eqref{eq:def_most_probable_tube}
 for a one-dimensional system
becomes a minimization problem in the $M$-dimensional space of coefficients.
 
To perform this minimization in practice, we use
the cma-es algorithm \cite{hansen_cma-espycma_2019}.
Unless stated otherwise, we use $M=40$ modes and
start the cma-es algorithm at the initial condition
$(c_1, c_2, ..., c_M) = (0,0, ..., 0)$
with an initial variance $\sigma_0 = 0.5$.
We run each minimization five times, and choose as MPT center
 the lowest of the five minima.

\end{document}